\documentstyle[11pt,amsfonts,fullpage]{amsart}

\def\endproof{\qed \medskip}
\def\blacksquare{\hbox to .60em {\vrule width .60em height .60em}}

\newtheorem{theorem}{Theorem}[section]

\newtheorem{corollary}[theorem]{Corollary}

\newtheorem{proposition}[theorem]{Proposition}

\newtheorem{remark}[theorem]{Remark}

\newtheorem{examples}[theorem]{Examples}

\begin{document}

\title[]{Geometric aspects of the AdS/CFT Correspondence}

\author[]{Michael T. Anderson}

\thanks{Partially supported by NSF Grant DMS 0305865}

\maketitle

\abstract
We discuss classical gravitational aspects of the AdS/CFT correspondence, 
with the aim of obtaining a rigorous (mathematical) understanding of 
the semi-classical limit of the gravitational partition function. The paper 
surveys recent progress in the area, together with a selection of new results 
and open problems. 
\endabstract

\tableofcontents

\setcounter{section}{-1}

\section{Introduction}
\setcounter{equation}{0}

 In this paper we discuss certain geometrical aspects of the AdS/CFT 
or Maldacena correspondence [42], [31], [47]. From a physics point of view, 
only the purely classical gravitational aspects of the correspondence on the 
AdS side are considered; thus no scalar, $p$-form or gauge fields, and no 
supergravity or string corrections are considered. On the CFT side, we only 
take account of the conformal structure on the boundary, and its corresponding 
stress-energy tensor. The discussion is also confined, by and large, to the 
Euclidean (or Riemannian) version of the correspondence. On the other hand, 
within this modest and restricted framework, there is quite a bit that is now 
known on a rigorous mathematical basis. 

 Broadly speaking, the AdS/CFT correspondence states the existence of a 
duality equivalence between gravitational theories (such as string or $M$ 
theory) on anti-de Sitter spaces $M$ and conformal field theories on 
the boundary at conformal infinity $\partial M$. In the restricted 
semi-classical framework above, the correspondence as formulated by Witten 
[47] states that 
\begin{equation} \label{e0.1}
Z_{CFT}[\gamma ] = \sum e^{-I(g)}, 
\end{equation}
where $Z_{CFT}$ is the partition function of a CFT attached to a 
conformal structure $[\gamma]$ on $\partial M$, and $I(g)$ is the 
renormalized Einstein-Hilbert action of an Einstein metric $g$ on $M$ 
with conformal infinity $[\gamma]$. The sum is over all manifolds and 
metrics $(M, g)$ with the given boundary data $(\partial M, [\gamma])$.

\medskip

  The main focus of the paper is on developing a framework in which the 
right side of (0.1) can be given a rigorous understanding. We survey 
existing work on geometrical aspects of the correspondence related to this 
issue, and discuss several new results, mostly in the later sections. 
Section 1 discusses general background information on the structure of 
conformally compact Einstein metrics, while \S 2 and \S 3 discuss uniqueness 
and existence issues for the Dirichlet-Einstein problem respectively. Section 
4 explains the role of positive scalar curvature boundary data in the existence 
theory. In \S 5 it is shown that the correspondence becomes much more explicit 
in the case of self-dual or anti-self-dual metrics on 4-manifolds. Finally 
in \S 6 we discuss the continuation to de Sitter-type metrics, and the 
construction of globally self-similar solutions to the vacuum Einstein 
equations. Throughout the text, a number of open questions and problems are 
presented.

\section{Conformally compact Einstein metrics}
\setcounter{equation}{0}

 Let $M$ be the interior of a compact $n+1$ dimensional manifold $\bar 
M$ with non-empty boundary $\partial M.$ A complete Riemannian metric 
$g$ on $M$ is $C^{m,\alpha}$ conformally compact if there is a defining 
function $\rho $ on $\bar M$ such that the conformally equivalent metric
\begin{equation} \label{e1.1}
\widetilde g = \rho^{2}g 
\end{equation}
extends to a $C^{m,\alpha}$ metric on the compactification $\bar M$. A 
defining function $\rho $ is a smooth, non-negative function on $\bar 
M$ with $\rho^{-1}(0) = \partial M$ and $d\rho  \neq $ 0 on $\partial 
M.$ The induced metric $\gamma  = \widetilde g|_{\partial M}$ is called 
the boundary metric associated to the compactification $\widetilde g.$ 
There are many possible defining functions, and hence many conformal 
compactifications of a given metric $g$, and so only the conformal class 
$[\gamma]$ of $\gamma$ on $\partial M$, called conformal infinity, is 
uniquely determined by $(M, g)$.  Any manifold $M$ carries many 
conformally compact metrics but we are mainly interested in Einstein 
metrics $g$, normalized so that
\begin{equation} \label{e1.2}
Ric_{g} = -ng. 
\end{equation}
It is well-known, and easily seen, that $C^{2}$ conformally compact 
Einstein metrics are asymptotically hyperbolic (AH), in that $|K_{g}+1| 
= O(\rho^{2}),$ where $K_{g}$ denotes sectional curvature of $(M, g)$, 
and these two notions will be used interchangeably. 

 A compactification $\bar g = \rho^{2}g$ as in (1.1) is called geodesic 
if $\rho (x) = dist_{\bar g}(x, \partial M).$ These compactifications 
are especially useful for computational purposes, and for the remainder 
of the paper we work only with geodesic compactifications. Each choice 
of boundary metric $\gamma  \in  [\gamma]$ determines a unique 
geodesic defining function $\rho$ associated to $(M, g)$. 

 The Gauss Lemma gives the splitting
\begin{equation} \label{e1.3}
\bar g = d\rho^{2} + g_{\rho}, \ \ g = \rho^{-2}(d\rho^{2} + g_{\rho}), 
\end{equation}
where $g_{\rho}$ is a curve of metrics on $\partial M.$ The 
Fefferman-Graham expansion [27] is a formal Taylor-type series expansion for 
the curve $g_{\rho}.$ The exact form of the expansion depends on 
whether $n$ is odd or even. For $n$ odd, one has 
\begin{equation} \label{e1.4}
g_{\rho} \sim  g_{(0)} + \rho^{2}g_{(2)} + .... + \rho^{n-1}g_{(n-1)} + 
\rho^{n}g_{(n)} + \rho^{n+1}g_{(n+1)} + ... 
\end{equation}
This expansion is even in powers of $\rho$ up to order $n-1$. The 
coefficients $g_{(2k)}$, $k \leq  (n-1)/2$ are locally determined by the 
boundary metric $\gamma  = g_{(0)};$ they are explicitly computable 
expressions in the curvature of $\gamma $ and its covariant 
derivatives. The term $g_{(n)}$ is transverse-traceless, i.e.
\begin{equation} \label{e1.5}
tr_{\gamma}g_{(n)} = 0, \ \  \delta_{\gamma}g_{(n)} = 0, 
\end{equation}
but is otherwise undetermined by $\gamma$; it depends on global aspects 
of the AH Einstein metric $(M, g)$. If $n$ is even, one has
\begin{equation} \label{e1.6}
g_{\rho} \sim  g_{(0)} + \rho^{2}g_{(2)} + .... + \rho^{n-2}g_{(n-2)} + 
\rho^{n}g_{(n)} + \rho^{n}\log \rho \ h + \rho^{n+1}g_{(n+1)} + ... 
\end{equation}
Again the terms $g_{(2k)}$ up to order $n-2$ are explicitly computable 
from the boundary metric $\gamma ,$ as is the transverse-traceless 
coefficient $h$ of the first $\log \rho $ term. The term $h$ is an 
important term relating to the CFT on $\partial M$; it is the metric 
variation of the conformal anomaly, cf. [22]. In odd dimensions, this 
always vanishes. The term $g_{(n)}$ satisfies 
\begin{equation} \label{e1.7}
tr_{\gamma}g_{(n)} = \tau , \ \  \delta_{\gamma}g_{(n)} = \delta ,  
\end{equation}
where $\tau $ and $\delta $ are explicitly determined by the boundary 
metric $\gamma $ and its derivatives, but $g_{(n)}$ is otherwise 
undetermined by $\gamma$ and as above depends on the global geometry of 
$(M, g)$. In addition $(\log\ \rho )^{k}$ terms appear at order $> n$ when 
$h \neq 0$. Note also that these expansions depend on the choice of 
boundary metric $\gamma \in [\gamma]$. Transformation properties of the 
coefficients $g_{(i)}$, $i \leq  n$,  and $h$ as $\gamma$ varies over 
$[\gamma]$ are also readily computable, cf. [22].
\medskip

 Mathematically, these expansions are formal, obtained by 
compactifiying the Einstein equations and taking iterated Lie 
derivatives of $\bar g$ at $\rho = 0$;
\begin{equation} \label{e1.8}
g_{(k)} = \frac{1}{k!}{\cal L}_{T}^{(k)}\bar g, 
\end{equation}
where $T = \bar \nabla\rho$. If $\bar g \in  C^{m,\alpha}(\bar M),$ then 
the expansions hold up to order $m+\alpha .$ However, boundary 
regularity results are needed to ensure that if an AH Einstein metric 
$g$ with boundary metric $\gamma $ satisfies $\gamma  \in  
C^{m,\alpha}(\partial M),$ then the compactification $\bar g \in  
C^{m,\alpha}(\bar M).$ 

 In both cases $n$ even or odd, the Einstein equations determine all higher order 
coefficients $g_{(k)}$, $k > n$, in terms of $g_{(0)}$ and $g_{(n)}$, so that an 
AH Einstein metric is formally determined by $g_{(0)}$ and $g_{(n)}$. 
The term $g_{(0)}$ corresponds to Dirichlet boundary data on $\partial 
M,$ while $g_{(n)}$ corresponds to Neumann boundary data, (in analogy 
with the scalar Laplace operator). Thus, on AH Einstein metrics, the 
correspondence
\begin{equation} \label{e1.9}
g_{(0)} \rightarrow  g_{(n)} 
\end{equation}
is analogous to the Dirichlet-to-Neumann map for harmonic functions. 
As discussed below, the term $g_{(n)}$ corresponds essentially to the 
stress-energy tensor of the dual CFT on $\partial M$. However, the map 
(1.9) is only well-defined per se if there is a unique AH Einstein metric with 
boundary data $\gamma  = g_{(0)}$. 

\medskip

 Understanding the correspondence (1.9) is one of the key issues in the 
AdS/CFT correspondence, restricted to the setting of the 
Introduction. Formally, knowing $g_{(0)}$ and $g_{(n)}$ allows one to 
construct the bulk gravitational field, that is the AH Einstein metric via 
the expansion (1.4) or (1.6). However, one needs to know that the 
expansions (1.4) or (1.6) actually converge to $g_{\rho}$ for this to 
be of any use. 

 More significantly, if $n$ is odd, given any real-analytic metric $g_{(0)}$ 
and symmetric bilinear form $g_{(n)}$ on $\partial M$, satisfying 
(1.5), there exists a unique $C^{\omega}$ conformally compact Einstein 
metric $g$ defined in a thickening $\partial M\times [0,\varepsilon)$ of 
$\partial M$. In particular, the expansion (1.4) converges to 
$g_{\rho}$. A similar statement holds when $n$ is even, cf. [6] for $n 
= 3$ and [39] or [45] for general $n$. Thus, the terms $g_{(0)}$ and 
$g_{(n)}$ may be specified arbitrarily and independently of each other, 
subject only to the constraint (1.5) or (1.7), to give ``local'' AH 
Einstein metrics. This illustrates the global nature of the correspondence (1.9).

\medskip

 Next, we turn to the structure of the moduli space of AH Einstein 
metrics on a given $(n+1)$-manifold $M$. Let $E = E^{\infty}$ be the 
space of AH Einstein metrics on $M$ which admit a $C^{\infty}$ 
compactification $\bar g$ as in (1.1). When $n$ is even, we assume here 
that $C^{\infty}$ means $C^{\infty}$ polyhomogeneous, i.e. $g_{\rho} = 
\phi (\rho , \rho^{n}\log \rho)$, where $\phi$ is a $C^{\infty}$ 
function of the two variables. The topology on $E$ is the $C^{\infty}$ 
(polyhomogeneous) topology on metrics on $\bar M$. Let ${\cal E}  = 
E/{\rm Diff}_{1}(\bar M)$, where ${\rm Diff}_{1}(\bar M)$ is the group of 
$C^{\infty}$ diffeomorphisms of $\bar M$ inducing the identity on 
$\partial M$, acting on $E$ in the usual way by pullback. (The CFT on 
$\partial M$ is a gauge-type theory, and so is diffeomorphism 
covariant, not diffeomorphism invariant; hence, it is natural to 
require diffeomorphisms fixing $\partial M$). 

  As boundary data, let $Met(\partial M) = Met^{\infty}(\partial M)$ be 
the space of $C^{\infty}$ metrics on $\partial M$ and 
${\cal C}  = {\cal C}(\partial M)$ the corresponding 
space of pointwise conformal classes. Occasionally we will also work 
with the spaces of real-analytic metrics $C^{\omega}$, or 
$C^{m,\alpha}$. 

 There is a natural boundary map, 
\begin{equation} \label{e1.10}
\Pi : {\cal E}  \rightarrow  {\cal C} , \ \ \Pi [g] = [\gamma], 
\end{equation}
which takes an AH Einstein metric $g$ on $M$ to its conformal infinity 
on $\partial M.$

 One then has the following general result on the structure of ${\cal 
E} $ and the map $\Pi ,$ building on previous work of Graham-Lee [30] 
and Biquard [13].
\begin{theorem} \label{t 1.1.}
 {\rm  [5,6]} Let $M$ be a compact, oriented $(n+1)$-manifold with boundary 
$\partial M.$ If ${\cal E} $ is non-empty, then ${\cal E} $ is a smooth 
infinite dimensional manifold. Further, the boundary map 
$$\Pi : {\cal E}  \rightarrow  {\cal C}  $$
is a $C^{\infty}$ smooth Fredholm map of index 0. Thus the derivative 
$D\Pi$ has finite dimensional kernel and cokernel, has closed range, and 
$$dim Ker D\Pi = dim Coker D\Pi.$$
\end{theorem}

 Implicit in Theorem 1.1 is the boundary regularity statement that an 
AH Einstein metric with $C^{\infty}$ conformal infinity has a 
$C^{\infty}$ (polyhomogeneous) geodesic compactification. When $n+1 =$ 
4, this boundary regularity has been proved in [4], [6], including 
the cases of $C^{\omega}$ and $C^{m,\alpha}$ regularity. In dimensions 
$n+1 > $ 4, boundary regularity has recently been proved by Chru\'sciel 
et al. [19] in the $C^{\infty}$ case. Moreover, when $\gamma\in C^{\omega}$, 
Kichenassamy [39] and Rendall [45] have proved that the expansions (1.4) 
and (1.6) converge to $g_{\rho}$.

\medskip

 In addition, the regular points of $\Pi$, that is the metrics in 
${\cal E}$ where $D\Pi$ is an isomorphism, are open and dense in 
${\cal E}$. Hence, if ${\cal E} \neq  \emptyset$, then $\Pi({\cal E})$ 
has non-empty interior in ${\cal C}$. Thus, if $M$ carries some 
AH Einstein metric, then it also carries a large set of them, 
parametrized at least by an open set in ${\cal C}$. The results above 
all hold with $E$ in place of ${\cal E}$, without essential changes. 

\medskip

 A basic issue in this area is the Dirichlet problem for AH Einstein 
metrics: given the topological data $(M, \partial M)$, and a conformal 
class $[\gamma ]\in{\cal C} ,$ does there exist a unique AH Einstein 
metric $g$ on $M$, with conformal infinity $[\gamma]$? In terms of the 
boundary map $\Pi$, global existence is equivalent to the surjectivity 
of $\Pi$, while uniqueness is equivalent to the injectivity of $\Pi$.

\medskip

 For Riemannian metrics, the Einstein-Hilbert action is (usually) given by 
\begin{equation} \label{e1.11}
I = -\frac{1}{16\pi G}\int_{M}(R-2\Lambda )dv - \frac{1}{8\pi 
G}\int_{\partial M}H dA,  
\end{equation}
where $R$ is the scalar curvature, $\Lambda$ the cosmological constant 
and $H$ is the mean curvature; (sometimes $I$ is replaced by $-I$). 
In the following, units are chosen so that $16\pi G = 1$. 

 Critical points of $I$ satisfy the Einstein equations
\begin{equation} \label{e1.12}
Ric - \frac{R}{2}g +\Lambda g = 0, 
\end{equation}
and in the normalization (1.2), $\Lambda  = \frac{1}{2}\frac{n-1}{n+1}R 
= -\frac{1}{2}n(n-1)$. However, both terms in (1.11) are infinite on 
metrics in ${\cal E} .$ A number of schemes have been proposed by 
physicists to obtain a finite expression for $I$ on ${\cal E} .$ 
Among these, the most natural is the holographic renormalization, 
c.f. [47], [34], [12], [22], described as follows. Given a fixed geodesic 
defining function $\rho$ for $g$, let $B(\varepsilon) = \{x\in (M, g): 
\rho (x) \geq  \varepsilon\}$. If $n$ is odd, from the expansion (1.4), 
one has an expansion of the volume of $B(\varepsilon)$ in the form
\begin{equation} \label{e1.13}
vol B(\varepsilon ) = v_{(0)}\varepsilon^{-n} + ... + 
v_{(n-1)}\varepsilon^{-1} + V + o(1), 
\end{equation}
where the terms $v_{(k)}$ are explicitly computable from $(\partial M, \gamma)$. 
For an Einstein metric as in (1.2), $R-2\Lambda = -2n$, so 
that 
\begin{equation} \label{e1.14}
-\int_{B(\varepsilon )}(R-2\Lambda )dv = 2n(v_{(0)}\varepsilon^{-n} + 
... + v_{(n-1)}\varepsilon^{-1} + V) + o(1). 
\end{equation}
A similar expansion of the boundary integral in (1.11) over 
$S(\varepsilon)$ has a form similar to (1.13), but with no constant 
term $V$. In fact local and covariant counterterms $v_{(k)}(\varepsilon)$ for 
the integral in (1.14), and the corresponding boundary integral, can be 
constructed in terms of the metric $\gamma_{\varepsilon}$ induced on the finite 
cut-off $S(\varepsilon) = \partial B(\varepsilon)$. These counterterms 
$v_{(k)}(\varepsilon)$, when suitably rescaled, converge to the counterterms 
$v_{(k)}$ at infinity; this is one important aspect of the AdS/CFT correspondence. 

  Thus, define the renormalized action $I^{ren}$ by
\begin{equation} \label{e1.15}
I^{ren}(g) = 2nV. 
\end{equation}
Similarly, if $n$ is even, the expansion (1.6) gives
\begin{equation} \label{e1.16}
vol B(\varepsilon ) = v_{(0)}\varepsilon^{-n} + ... + 
v_{(n-2)}\varepsilon^{-2} + L\log \varepsilon  + V + o(1), 
\end{equation}
and again the terms $v_{(k)}$ as well as $L$ are explicitly computable from 
$(\partial M, \gamma)$. The coefficient $L$, equal to the integral of $tr g_{(n)}$ 
over $\partial M$, agrees with the conformal anomaly of the dual CFT on $\partial M$ 
in all known cases, [34]. The renormalized action is again defined by (1.15).

\medskip

 When $n$ is odd, $I^{ren}$ is independent of the choice of boundary 
metric $\gamma\in [\gamma]$, and thus $I^{ren}$ is a smooth functional 
on the moduli space ${\cal E}$. When $n$ is even, this is not the 
case; $I^{ren}$ does depend on the choice of boundary metric, and so 
only gives a smooth functional on the space $E$, (or $E$ quotiented out 
by diffeomorphisms equal to the identity to order $n$ at $\partial M$). 
On the other hand, $L$ is independent of the choice of boundary metric. 

\medskip

 Consider the variation of $I^{ren}$ at a given $g\in E$, i.e.
\begin{equation} \label{e1.17}
dI^{ren}(\dot g) = \frac{d}{dt}I^{ren}(g+t\dot g), 
\end{equation}
where $\dot g$ is tangent to $E$. Thus, $dI^{ren}$ may be 
considered as a 1-form on the manifold $E$ (or ${\cal E}$ when $n$ is 
odd). In general, the differential $dI^{ren}$ is the stress-energy (or 
energy-momentum) tensor of the action. Since Einstein metrics are critical 
points of $I$ or $I^{ren}$, it is clear that $dI^{ren}$ must be supported on 
$\partial M$. In [22] it is proved that
\begin{equation} \label{e1.18}
dI^{ren} = -\frac{n}{2}g_{(n)} + r_{(n)}, 
\end{equation}
where $r_{(n)} =$ 0 if $n$ is odd, and is explicitly determined by 
$\gamma  = g_{(0)}$ if $n$ is even. Thus
\begin{equation} \label{e1.19}
dI^{ren}(\dot g) = -\frac{n}{2}\int_{\partial M} \langle g_{(n)} + r_{(n)}, 
\dot g_{(0)} \rangle dv_{\gamma}, 
\end{equation}
where $\dot g_{(0)}$ is the variation of the boundary metric $\gamma$ 
induced by $\dot g$. On the gravitational side, the 1-form $dI^{ren}$ is the 
(Brown-York) quasi-local stress-energy tensor $T$; via the AdS/CFT 
correspondence, this corresponds to the expectation value $\langle T \rangle$ 
of the stress-energy tensor of the dual CFT on $\partial M$.

\medskip

 In dimension 4, the renormalized action or volume can be given a quite 
different interpretation. Namely, by means of the Chern-Gauss-Bonnet 
theorem, one finds, on $(M^{4}, g)$,
\begin{equation} \label{e1.20}
\int_{M}|W|^{2} = 8\pi^{2}\chi (M) - I^{ren}, 
\end{equation}
where $\chi (M)$ is the Euler characteristic and $W$ is the Weyl 
curvature of $(M, g)$, cf. [3]. Thus the renormalized action, which 
involves only the scalar curvature and volume, in fact controls much 
more; it controls $L^{2}$ norm of the Weyl curvature $W$ on-shell, i.e. 
on ${\cal E} .$ 

 Since the left side of (1.20) is non-negative, an immediate 
consequence is that the renormalized action is uniformly bounded above 
on the full space ${\cal E}$, depending only on a lower bound for 
$\chi (M)$. Moreover, $I^{ren}$ has an absolute maximum on hyperbolic 
metrics. Thus for example on the 4-ball $B^{4},$ the Poincar\'e metric 
has the largest action among all AH Einstein metrics on $B^{4}.$ It is 
an interesting open question whether such a result also holds in higher 
dimensions.

\section{Uniqueness Issue.}
\setcounter{equation}{0}

 It is not unreasonable to believe that there should be some relation 
between the existence and uniqueness problems for the 
Einstein-Dirichlet problem. For example, the usual Fredholm alternative 
relates these two issues at the linearized level. In this section, we 
discuss the uniqueness question on the basis of a selection of 
examples. 

\begin{examples} \label{e 2.1}
 {\rm The first example of non-uniqueness was that found by 
Hawking-Page [32] in their analysis of the AdS Schwarzschild, or AdS 
$S^{2}$ black hole metric. In general dimensions, this is a curve of AH 
Einstein metrics on $M = {\Bbb R}^{2}\times S^{n-1}$ given by
\begin{equation} \label{e2.1}
g_{m} = V^{-1}dr^{2} + Vd\theta^{2} + r^{2}g_{S^{n-1}(1)}, 
\end{equation}
where 
\begin{equation} \label{e2.2}
V(r) = 1+r^{2}-\frac{2m}{r^{n-2}}. 
\end{equation}
Here $r \in  [r_{+},\infty )$, where $r_{+}$ is the largest root of $V$, 
and the circular parameter $\theta  \in  [0,\beta]$, where $\beta  = 
4\pi r_{+}/(nr_{+}^{2}+(n-2))$. It is easy to see that the conformal 
infinity of $g_{m}$ is given by the conformal class of the product 
metric on $S^{1}(\beta )\times S^{n-1}(1)$. The action of $g_{m}$ is given by
\begin{equation} \label{e2.3}
I^{ren}(g_{m}) = -\beta\omega_{n-1}(r_{+}^{n} - r_{+}^{n-2} + c_{n}), 
\end{equation}
where $c_{n} =$ 0 if $n$ is odd, and $c_{n} = 
(-1)^{n/2}\frac{(n-1)!!^{2}}{n!}$ if $n$ is even, with $\omega_{n-1} = 
vol S^{n-1}(1)$, cf. [24]. The stress-energy tensor of $g_{m}$ is
\begin{equation} \label{e2.4}
dI^{ren} = -(r_{+}^{n} + r_{+}^{n-2} + \frac{2c_{n}}{n-1})diag(1-n, 1, 
...1), 
\end{equation}
cf. [25]. As a function of $m \in  (0,\infty ),$ observe that $\beta $ 
has a maximum value of $\beta_{0} = 2\pi /(n(n-2))^{1/2}$, and 
for every $m \neq  m_{0},$ there are two values $m^{\pm}$ of $m$ giving 
the same value of $\beta$. Thus two metrics have the same conformal 
infinity; the boundary map $\Pi$ is a fold map, (of the form $x 
\rightarrow  x^{2}$) along the curve $g_{m}$. Exactly the same formulas 
and behavior hold if $S^{n-1}(1)$ is replaced by any closed Einstein 
manifold $(N, g_{N})$ with $Ric_{g_{N}} = (n-2)g_{N},$ with 
$\omega_{n-1}$ replaced by $vol_{g_{N}}N.$

 Note that if one allows the filling manifold $M$ to change, a further 
metric has the same conformal infinity. Thus, choose $M = B^{n+1}/{\Bbb 
Z}  = S^{1}\times {\Bbb R}^{n},$ with the quotient of the hyperbolic metric 
$g_{-1}$ on $B^{n+1}$ by a translation isometry along a geodesic. }
\end{examples}

\begin{examples} \label{e 2.2}

{\rm  As discussed in [4], a more drastic example of non-uniqueness 
occurs in the family of AdS toral black hole metrics. These are metrics 
on $M = {\Bbb R}^{2}\times T^{n-1},$ where $T^{n-1}$ is the $(n-1)$-torus, 
and the standard form of the metrics $g_{m}$ is the same as in (2.1), 
with $S^{n-1}(1)$ replaced by any flat metric on $T^{n-1}$ and $V$ in 
(2.2) replaced by $V(r) = r^{2} - \frac{2m}{r^{n-2}}$, $\beta  = 4\pi 
/nr_{+}$. The conformal infinity of these metrics is the flat metric on 
the product $S^{1}(\beta )\times T^{n-1}$. Here $\beta$ is monotone in $m$, 
and so on this space of metrics, the boundary map $\Pi$ is 1-1. The 
action and stress-energy tensor are given by, [24], [25]: 
\begin{equation} \label{e2.5}
I^{ren}(g_{m}) = -\beta\omega_{n-1}r_{+}^{n} , \ \  dI^{ren} = 
-r_{+}^{n}diag(1-n, 1, ..., 1), 
\end{equation}
where $\omega_{n-1} = vol T^{n-1}.$

 However, the actual situation is a little more subtle. The metrics 
$g_{m}$ are all locally isometric, and so are isometric in the 
universal cover ${\Bbb R}^{2}\times {\Bbb R}^{n-1},$
\begin{equation} \label{e2.6}
\widetilde g_{m} = V^{-1}dr^{2} + Vd\theta^{2} + r^{2}g_{{\Bbb 
R}^{n-1}}. 
\end{equation}
Let $(T^{n}, g_{0})$ be any flat metric on the $n$-torus, and let $\sigma$ 
be any simple closed geodesic in $(T^{n}, \sigma ).$ Topologically, 
one may glue on a disc $D^{2} = {\Bbb R}^{2}$ onto $\sigma$ to obtain 
a solid torus ${\Bbb R}^{2}\times T^{n-1}$. Metrically, this is carried out 
as follows. Given $\sigma$, for any $R$ sufficiently large, there 
exists $S$, and a covering map $\pi  = \pi_{R}: S^{1}\times {\Bbb R}^{n-1} 
\rightarrow  T^{n}$, such that 
$$\pi (S^{1}) = \sigma , \ \ {\rm and} \ \  \pi^{*}(S^{2}g_{0}) = V(R)d\theta^{2} + 
R^{2}g_{{\Bbb R}^{n-1}}. $$
Here $S$ is determined by the relation that $V(R)^{1/2}\beta  = 
SL(\sigma)$, where $L(\sigma)$ is the length of $\sigma$ in $(T^{n}, g_{0})$. 
Thus $\pi$ takes the circle factor $S^{1}$ to $\sigma$ and maps the flat metric 
on $S^{1}\times {\Bbb R}^{n-1}$ to $S^{2}g_{0}$ on $T^{n}$. The map $\pi$ is 
given by dividing by a unique (twisted) isometric ${\Bbb Z}^{n-1}$-action 
on $S^{1}\times T^{n-1}$ and this action clearly extends to an isometric 
${\Bbb Z}^{n-1}$-action on $\widetilde g_{m}$. Letting $R \rightarrow \infty$ 
and taking the corresponding limiting map $\pi$ and ${\Bbb Z}^{n-1}$-action 
gives a ``twisted'' toral black hole metric
\begin{equation} \label{e2.7}
\hat g_{m} = V^{-1}dr^{2} + [Vd\theta^{2} + r^{2}g_{{\Bbb 
R}^{n-1}}]/{\Bbb Z}^{n-1} 
\end{equation}
on ${\Bbb R}^{2}\times T^{n-1}$ with conformal infinity $(T^{n}, g_{0})$. 

 By varying the choices of $\sigma$, this gives infinitely many isometrically 
distinct AH Einstein metrics on ${\Bbb R}^{2}\times T^{n-1}$ with the same 
conformal infinity $(T^{n}, \sigma)$, so that $\Pi$ is $\infty$-to-1; (note 
however that these metrics are all locally isometric). These metrics all 
lie in distinct components of the moduli space ${\cal E}$, so ${\cal E}$ has 
infinitely many components on ${\Bbb R}^{2}\times T^{n-1}$; these components 
are permuted by the action of ``large'' diffeomorphisms on the boundary $T^{n}$, 
not isotopic to the identity, corresponding to the choices of simple closed geodesic 
$\sigma$. 

\medskip

 One may take limits of any infinite sequence of these metrics, with 
fixed conformal infinity, by taking $L(\sigma ) \rightarrow  \infty .$ 
All sequences have a unique limit given by the hyperbolic cusp metric
\begin{equation} \label{e2.8}  
g_{C} = ds^{2} + e^{2s}g_{0}, 
\end{equation}
on ${\Bbb R} \times T^{n}.$ It is not difficult to compute exactly the action 
$I^{ren}$ of $\hat g_{m},$ and it is easy to see that as $L(\sigma ) 
\rightarrow  \infty ,$ 
$$I^{ren}(\hat g_{m}) \rightarrow  I^{ren}(g_{C}) = 0, $$
(corresponding to $\beta  \rightarrow  \infty$). }
\end{examples}

\begin{remark} \label{r 2.3.}
{\rm  Let $(N, g_{N})$ be any closed $(n-1)$-dimensional Einstein 
manifold, with $Ric_{g_{N}} = -(n-1)g_{N}.$ Such metrics generate AdS 
black hole metrics just as in (2.1); thus 
\begin{equation} \label{e2.9}
g_{m} = V^{-1}dr^{2} + Vd\theta^{2} + r^{2}g_{N}, 
\end{equation}
is an AH Einstein metric on ${\Bbb R}^{2}\times N,$ with conformal infinity 
$S^{1}(\beta )\times (N,g_{N}).$ Here $V = -1+r^{2} - 2m/r^{n-2}$, $r_{+} > 
0$ is the largest root of $V$ and $\beta  = 4\pi r_{+}/(nr_{+}^{2}-(n-2))$. 
Again $\beta$ is a monotone function of $m$. 
However, $g_{m}$ is well-defined for negative values of $m$; in fact, 
$g_{m}$ is well-defined for $m \in  [m_{-}, \infty)$, where 
\begin{equation} \label{e2.10}
m_{-} = -\frac{1}{n-2}(\frac{n-2}{n})^{n/2}, \ \ {\rm with} \ \  r_{+} = 
(\frac{n-2}{n})^{1/2}. 
\end{equation}
For the extremal value $m_{-}$ of $m$, $V(r_{+}) = V' (r_{+}) = 0$, and a 
simple calculation shows that the horizon $\{r = r_{+}\}$ occurs at 
infinite (geodesic) distance to any given point in $(M, g_{m_{-}})$; 
the horizon in this case is called degenerate, (with zero surface 
gravity). Note that $\beta (m_{-}) = \infty$, so that the $\theta$-circles 
are in fact lines ${\Bbb R}$. As $m$ decreases to $m_{-}$, 
the horizon diverges to infinity, (in the opposite direction from the 
conformal infinity), while the length of the $\theta$-circles expands 
to $\infty$. Thus, the metric $g_{m_{-}}$ is a complete metric on the 
manifold ${\Bbb R} \times {\Bbb R} \times N = {\Bbb R}^{2}\times N$, but 
is no longer conformally compact; the conformal infinity is ${\Bbb R} \times (N, g_{N})$. 
The action also diverges to $-\infty $ as $m \rightarrow  m_{-}$.

 However, one may divide the infinite $\theta$-factor of the extremal 
metric $g_{m_{-}}$ by ${\Bbb Z}$ to obtain a complete metric $\hat g$ 
on ${\Bbb R} \times S^{1}\times N$. The metric $\hat g$ is an AH Einstein metric 
with a single cusp-like end, and with conformal infinity $S^{1}\times (N, 
g_{N})$; the length of the $S^{1}$ factor may be arbitrary. This is a 
non-standard example of an AH Einstein metric, with a cusp-like end, as 
opposed to the standard hyperbolic cusp of (2.8). 

 Note however that in contrast to the situation in Examples 2.2, $\hat g$ 
does not arise as a limit of the curve $g_{m}$ as $m \rightarrow  m_{-}$; 
as $m \rightarrow  m_{-},$ the conformal infinity of $g_{m}$ also degenerates. }
\end{remark}

\begin{examples} \label{ex2.4}

{\rm  As a final example of non-uniqueness, let $(N, g_{0})$ be any 
complete, geometrically finite hyperbolic $(n+1)$-manifold. We assume 
that $N$ has a conformal infinity $(\partial N, \gamma_{0}),$ as well 
as a finite number of parabolic or cusp ends, of the form (2.8). There 
are numerous examples of such manifolds in any dimension. If $n$ is 
odd, the renormalized action and stress-energy tensor are given by
\begin{equation} \label{e2.11}
I^{ren}(N, g_{0}) = (-1)^{m}2n\frac{2^{2m}\pi^{m}m!}{(2m)!}\chi (N), \ \ 
dI^{ren}(N, g_{0}) = 0, 
\end{equation}
where $n = 2m-1$, cf. [26]. If $n$ is even, an explicit general formula for 
the renormalized action is not known, although of course it is finite, while 
$dI^{ren}(N, g_{0})$ is explicitly computable from the $g_{(2)}$ term in 
(1.6), cf. [22] for example. 

\medskip

 Now one may truncate and cap off the cusp ends of $(N, g)$ by glueing in 
solid tori ${\Bbb R}^{2}\times T^{n-1}$ with boundary $\partial ({\Bbb 
R}^{2}\times T^{n-1}) = T^{n}.$ In 3 dimensions, this is the process of 
hyperbolic Dehn filling, due to Thurston. Essentially exactly as in 
Examples 2.2, a disc ${\Bbb R}^{2} = D^{2}$ can be attached to any 
simple closed geodesic $\sigma$ in $T^{n}.$

 Using the Dehn filling results in [7], G. Craig has recently shown [21] 
that all the cusp ends of $N$ may be capped off in this way to 
produce infinitely many distinct manifolds $M_{i},$ with AH Einstein 
metrics $g_{i},$ all with conformal infinity given by the original 
$(\partial N, \gamma_{0})$. The construction implies that as the lengths 
of all geodesics $\sigma_{i}$ diverge to infinity, $(M_{i}, g_{i})$ 
converges to the original manifold $(N, g_{0})$ in the Gromov-Hausdorff 
topology. Moreover, for all $i$ large, $I^{ren}(M_{i}, g_{i}) < 
I^{ren}(N, g_{0})$, and 
\begin{equation} \label{e2.12}
I^{ren}(M_{i}, g_{i}) \rightarrow  I^{ren}(N, g_{0}),  \ \ \    
dI^{ren}(M_{i}, g_{i}) \rightarrow  dI^{ren}(N, g_{0}), 
\end{equation}
in any dimension. }
\end{examples}

 Taken together, the results above suggest that in general, there may be some 
difficulties in obtaining a well-defined (purely gravitational) semi-classical 
partition function $Z_{AdS}$. Analogous difficulties in defining the partition 
function for Euclidean quantum gravity have been discussed briefly in [9]. Thus, 
given $(\partial M, \gamma)$, the correspondence (0.1) requires summing 
over the moduli space of all AH Einstein manifolds $(M, g)$ with the given 
boundary data $(\partial M, \gamma)$. In the infinite sets of AH 
Einstein metrics with fixed boundary data constructed above, the action 
$I^{ren}$ is uniformly bounded above and converges to a limit. 
Moreover, all metrics are strictly stable, in the sense that the 
$2^{\rm nd}$ variation of the action among transverse-traceless perturbations 
vanishing at infinity is positive definite - there are no negative or 
zero eigenmodes present. Hence, all solutions contribute a definite 
positive amount to the partition function, and so the partition 
function is likely to be badly divergent. In slightly more detail, the 
zero-loop approximation to the partition function is badly divergent, 
while the one-loop appoximation is also likely to be, unless there 
happen to be infinitely many other AH Einstein metrics giving rise to 
cancellations.

 As will be seen below, this phenomenon does not occur when the 
boundary metric $\gamma $ has positive scalar curvature, at least in 
dimension 4. This is perhaps then another reason for restricting the 
correspondence to boundary data of positive scalar curvature, as 
suggested by Witten [47], [48] for reasons related to the stability of 
the CFT. 

 However, a simple sum as in (0.1), even with the addition of higher 
loop corrections, may be ignoring certain important geometric 
information. Thus, suppose one had a finite dimensional connected 
moduli space $\Lambda$ of solutions with fixed boundary data, (so that 
there is a finite dimensional space of zero modes). In this case, one 
would not simply sum over the distinct solutions $g_{\lambda}$, but 
integrate the function $e^{-I}$ over the (presumably finite volume) moduli space 
$\Lambda$ with respect to the volume form induced by the $L^{2}$ metric 
on the space of metrics. It seems reasonable and natural that a similar 
prescription should be used when one has an infinite sequence of isolated 
points $\{g_{i}\}$ converging to a limit set $X_{\infty}$. The metrics 
$g_{i}$ satisfy
$$dist_{L^{2}}(x_{i}, X_{\infty}) \rightarrow 0, \ \ {\rm as} \ \ 
i \rightarrow \infty,$$ 
and it is natural to include weight factors, depending on 
$dist_{L^{2}}(x_{i}, X_{\infty})$, in the sum (0.1). What is not clear is 
exactly what weight factors one should choose.

 Such infinite behavior in the gravitational partition function does 
appear in the remarkable paper of Dijkgraff et al. [23], where the 
authors deal with the infinite family of BTZ black hole metrics, 
parametrized by relatively prime integers $(c,d)$ corresponding to simple 
closed geodesics on $T^{2}$; in the terminology above, these are 
just the different hyperbolic Dehn fillings of a 2-torus. The partition 
function found in [23] does have suitable weight factors, leading to a 
convergent sum.

\section{Existence Issue.}
\setcounter{equation}{0}

 Next we turn to the global existence question, i.e. the surjectivity 
of the boundary map $\Pi$. Locally, the map $\Pi$ is quite simple; 
its domain and target are smooth manifolds and the linearization $D\Pi 
$ has finite equi-dimensional kernel and cokernel. However, globally 
the domain ${\cal E}$ of $\Pi$ is highly non-compact. To obtain a 
good global theory relating the domain and image of $\Pi$, one needs 
the map $\Pi$ to be proper, i.e. for any compact set $K \subset {\cal C}$, 
$\Pi^{-1}(K)$ is compact in ${\cal E}$. In particular, for any 
$[\gamma ]\in{\cal C}$, $\Pi^{-1}([\gamma])$ should be a compact set in 
${\cal E}$. If this fails, for instance if $\Pi^{-1}([\gamma])$ is 
not compact, one needs to understand the possible limit structures of 
metrics in $\Pi^{-1}([\gamma])$.

 The lack of uniqueness or even finiteness discussed in \S 2 shows that in 
general $\Pi$ is not proper. Any general results on the compactness of a space of 
Einstein metrics having a compact set of boundary metrics must rely on 
a simpler theory of compactness of Einstein metrics on closed 
manifolds, i.e. the study of moduli spaces of Einstein metrics on 
closed manifolds. In dimension 2, the moduli space of Einstein metrics 
is described by Teichm\"uller theory. Unfortunately, in general 
dimensions, such a theory does not exist, and seems out of current 
reach. However, there is quite a well-developed theory of moduli of 
Einstein metrics on closed manifolds in 4-dimensions, and this allows 
one to develop an analogous theory in the case of AH Einstein metrics.

 The results described below are thus restricted to 4-manifolds $M$, with 
$\partial M$ a 3-manifold. It seems reasonable that these results can 
be generalized to higher dimensions in the presence of extra symmetry 
via Kaluza-Klein type symmetry reductions, and progress in this 
direction would be very interesting.

\medskip

 Let ${\cal C}^{0}$ be the space of conformal classes on a $3$-manifold 
$\partial M$ which have a {\it  non-flat}  representative metric of 
non-negative scalar curvature. (Of course not all 3-manifolds admit 
such a metric). Let ${\cal E}^{0} = \Pi^{-1}({\cal C}^{0}),$ and 
consider the restricted map
\begin{equation} \label{e3.1}
\Pi^{0}: {\cal E}^{0} \rightarrow  {\cal C}^{0}. 
\end{equation}

\begin{theorem} \label{t 3.1.}
 {\rm [5]} Let $M$ be a 4-manifold satisfying
\begin{equation} \label{e3.2}
H_{2}(\partial M) \rightarrow  H_{2}(M) \rightarrow  0. 
\end{equation}
Then the map $\Pi^{0}$ in (3.1) is proper. Further, $\Pi $ has a 
well-defined degree, given by
\begin{equation} \label{e3.3}
deg \Pi  = \sum_{g_{i}\in\Pi^{-1}[\gamma ]}(-1)^{ind g_{i}}. 
\end{equation}
\end{theorem}
Here $[\gamma]$ is any regular value of $\Pi $ in ${\cal C}^{0}$, 
(recall the regular values are dense in ${\cal C}^{0}$). Since 
$\Pi^{0}$ is proper, the sum above is finite, and $ind_{g_{i}}$ is the 
$L^{2}$ index of $g_{i}$, that is the dimension of the space of 
transverse-traceless $L^{2}$ forms on which the $2^{\rm nd}$ variation 
of the action is negative definite, (the number of negative eigenmodes).

 It is obvious from the definition that 
\begin{equation} \label{e3.4}
deg \Pi  \neq  0 \Rightarrow  \Pi^{0} \ \ {\rm is \ surjective}. 
\end{equation}
On the other hand, $\Pi^{0}$ may or may not be surjective when 
$deg \Pi^{0} = 0$. Similarly, $deg \Pi^{0} = \pm 1$ does not imply 
uniqueness of an AH Einstein metric with a given conformal infinity; 
it implies that generic boundary metrics have an odd number of 
AH Einstein filling metrics.

\begin{remark} \label{r 3.2.}
 {\rm  The condition (3.2) is used only to rule out degeneration of 
Einstein metrics to orbifolds. If one enlarges the space ${\cal E}$ to 
include orbifold Einstein metrics ${\cal E}_{s},$ then $\Pi^{0}$ is 
proper on the enlarged space $\hat {\cal E} = {\cal E}  \cup  {\cal 
E}_{s}.$ However, it is not currently known if $\hat {\cal E}$ has the 
structure of a smooth manifold, or of a manifold off a set of 
codimension $k$, for some $k \geq 2$, although one certainly expects 
this to be the case. }
\end{remark}

 Theorem 3.1 implies in particular that there are only finitely many 
components $C_{\lambda}$ of the moduli space ${\cal E}^{0}$ for which 
$\cap_{\lambda}\Pi (C_{\lambda}) \neq  \emptyset$, i.e. only finitely 
many components have a given boundary metric in common. Of course the 
degree $deg \Pi^{0}$ may depend on the choice of the component. This 
result also holds when one allows the manifold $M$ to vary. Thus, given 
any $K <  \infty$, there are only finitely many diffeomorphism types 
of 4-manifolds $M$, with a given boundary $\partial M$, with $\chi (M) 
\leq K$, and for which $\Pi({\cal E})$ contains any given element 
$[\gamma]\in{\cal C}^{0}$. 

 Thus, the infinities discussed in \S 2 cannot arise when $R_{\gamma} > 0$, 
and so the sum in (0.1) is essentially well-defined. The sum could be 
infinite only if there exist $(M_{i}, g_{i})$ with $\Pi(g_{i}) = 
\gamma$, with $\chi(M_{i}) \rightarrow +\infty$, (slightly analogous 
to the divergence of the string partition function). The role of the 
hypothesis $R_{\gamma} >$ 0 will be explained in more detail in \S 4, 
but we outline the general idea of the proof of Theorem 3.1 to explain 
how this condition arises.

\medskip

 By way of background, consider first the structure of the moduli space 
 of unit volume Einstein metrics on a fixed closed 4-manifold. Modulo
the possibility of orbifold degenerations, the overall structure of the 
moduli space is quite similar to that of the Teichm\"uller theory for the 
moduli space of constant curvature metrics on a closed surface. Recall 
that sequences $\{g_{i}\}$ of unit area constant curvature metrics on a 
surface $S$ have subsequences that either: 

$\bullet$ Converge to a limit $(S, g)$, (for example on $S^{2}$ where 
the moduli space is a point), 

$\bullet $ Collapse in the sense that the injectivity radius converges 
to 0 everywhere, (as in the case of a divergent sequence of flat 
metrics on $T^{2}),$ 

$\bullet$ Form cusps $(N, g)$, $N \subset S$, as in the case of 
hyperbolic metrics on $S$. Thus $S$ is a finite union of hyperbolic 
surfaces with cusp ends $(N_{k}, g_{k})$ together with a finite number 
of annuli ${\Bbb R} \times S^{1}$ which are collapsed by the sequence 
$\{g_{i}\}.$

 A similar basic trichotomy holds in dimension 4, cf. [2]. Thus, 
analogous to (1.20), the Chern-Gauss-Bonnet theorem implies a uniform 
upper bound on the $L^{2}$ norm of the Weyl curvature of an Einstein 
metric on a 4-manifold $M$. Using this, sequences of such metrics have 
subsequences that either converge, collapse, or form cusps as above, 
although one must allow also for the formation of orbifold singularities.

\medskip

 Now suppose instead that $\{g_{i}\}$ is a sequence of AH Einstein 
metrics on a 4-manifold $M$ for which the corresponding conformal 
infinities $\gamma_{i}$ are contained in a compact set, so that 
\begin{equation} \label{e3.5}
\gamma_{i} \rightarrow  \gamma , 
\end{equation}
(in a subsequence). Then using (3.5) and (1.20), one can again prove 
$I^{ren}(g_{i})$ remains uniformly bounded, (so that, roughly speaking, 
$I^{ren}$ is proper on ${\cal C} )$ and the trichotomy above still 
holds. The possibility of collapse can also be ruled out by the control on 
the boundary metrics. However, in general, the formation of cusps 
cannot be ruled out, (as seen from the examples in \S 2).

 More precisely, define an AH Einstein metric with cusps $(N, g)$ to be a 
complete Einstein metric $g$ on an $(n+1)$-manifold $N$ which has two 
types of ends, namely AH ends and cusp ends. A cusp end of $(N, g)$ is an 
end $E$ such that $vol_{g}E < \infty ,$ and hence is not conformally 
compact. The bound on the volume follows from the bound on the 
renormalized action, via (1.15). On any divergent sequence of points 
$x_{k}\in E,$ the injectivity radius $inj_{g}(x_{k}) \rightarrow $ 0 as 
$k \rightarrow  \infty$, so the metric $g$ is collapsing at infinity 
in $E$. For instance, as discussed in Examples 2.2, infinite sequences of 
twisted toral black hole metrics limit on a complete hyperbolic cusp metric 
(2.8). Similarly, infinite sequences in Examples 2.4 limit on a complete 
hyperbolic manifold with cusp ends.

 It will be seen in \S 4 that the hypothesis $R_{\gamma} > 0$, (or 
$R_{\gamma} \geq  0$ and $\gamma$ not Ricci-flat), rules out the possible 
formation of cusps, (in any dimension). This shows that $\{g_{i}\}$ 
above has a subsequence converging to a limit AH Einstein metric $g$ on 
$M$, so that $\Pi^{0}$ is proper.

\medskip

 The examples of cusp formation discussed in \S 2 all take place on sequences 
of metrics $g_{i}$ lying either in distinct components of ${\cal E}$, or on 
different smooth manifolds $M_{i}$. Another interesting open question is whether 
cusps can actually form within a given or fixed component of ${\cal E}$, on a 
fixed manifold $M$. On closed manifolds, it is clear that cusps can form 
at the endpoints of curves of Einstein metrics. For example, let $M = 
\Sigma_{g_{1}}\times \Sigma_{g_{2}}$ be a product of surfaces of genus $g_{i} 
\geq $ 2. Products of hyperbolic metrics on each surface are Einstein 
metrics on $M$ and so as in Teichm\"uller theory there are smooth curves 
of Einstein metrics limiting on cusps $(N, g)$ associated to $M$. However, 
it is not so easy to see, and in any case is unknown, if analogues of such 
constructions hold in the AH setting; compare also with Remark 2.3.

 It would also be very interesting to know if the possible formation of 
cusps is restricted by the topology of the ambient manifold $M$. For 
example, the topological condition (3.2) rules out the formation of 
orbifold singularities. In the example above on 
$M = \Sigma_{g_{1}}\times \Sigma_{g_{2}}$, this is clearly the case; 
the fundamental group of the collapsed region is non-trivial and 
injects in the fundamental group of $M$. One might conjecture for 
instance that on the 4-ball $B^{4}$, or $(n+1)$-ball $B^{n+1}$, cusp 
formation is not possible.

  If one knows that no cusp formation is possible on limits of sequences in 
${\cal E}$ = ${\cal E}(M)$ within a compact set of boundary metrics, then 
Theorem 3.1 holds in general, without the restriction to ${\cal E}^{0}$. 

\medskip

 Consider briefly the situation in general where cusps may form. Given a fixed 
4-manifold $M$, let $\bar {\cal E}$ be the completion of ${\cal E} $ in the 
Gromov-Hausdorff topology. Thus $(X, g) \in \bar {\cal E}$ iff there is a sequence 
$\{g_{i}\}\in{\cal E} $ such that $(M, g_{i}) \rightarrow (X, g)$ in the (pointed) 
Gromov-Hausdorff topology. The analysis above implies that, in general, 
\begin{equation} \label{e3.6}
\bar {\cal E} = {\cal E}  \cup  {\cal E}_{s} \cup  {\cal E}_{c}, 
\end{equation}
where ${\cal E}_{s}$ consists of orbifold AH Einstein metrics and 
${\cal E}_{c}$ consists of AH Einstein metrics with cusps, associated 
to $M$. The boundary map $\Pi$ extends to a continuous map
\begin{equation} \label{e3.7}
\bar \Pi: \bar {\cal E} \rightarrow  {\cal C} , 
\end{equation}
and $\bar \Pi$ is proper. However, the structure of $\bar {\cal E}$ is 
not well understood. If for example $\bar {\cal E}$ is a manifold off a 
singular set of codimension at least 2, then $deg \bar \Pi$ is well 
defined, and is given by (3.3). In particular, (3.4) then holds. 
Moreover, one clearly has $deg \bar \Pi = deg \Pi^{0}$, when 
${\cal C}^{0} \neq  \emptyset$. However, if $\bar {\cal E}$ is something 
like a manifold with boundary, then $deg \bar \Pi$ is not well-defined 
and the global behavior of $\bar \Pi$ is less clear.

\medskip

 We now return to applications of Theorem 3.1 itself. The degree $deg \Pi^{0}$ 
is a smooth invariant of $(M, \partial M)$. In many specific cases, the degree 
can be calculated by means of the following isometry extension result; this result 
is very natural from the perspective of the AdS/CFT correspondence, and holds in all 
dimensions.

\begin{theorem} \label{t 3.3.}
  {\rm [5]} Let $g$ be a $C^{2}$ conformally compact Einstein metric 
on an $(n+1)$ manifold with conformal infinity $[\gamma]$ on $\partial M$. 
Then any 1-parameter group of conformal isometries of 
$(\partial M, \gamma)$ extends to a 1-parameter group of isometries 
of $(M, g)$.
\end{theorem}

 In particular, any AH Einstein metric whose conformal infinity is 
highly symmetric is itself highly symmetric. Einstein metrics which 
have a transitive or cohomogeneity 1 isometric group action have 
essentially been classified. Using this, the degree $deg \Pi^{0}$ can 
be computed in a number of interesting cases; the currently known 
results are summarized in the table below.

\bigskip

\begin{center}
Table
\end{center}

\renewcommand{\arraystretch}{1.5}

\begin{center}

\begin{tabular}{|l|l|l|l|l|}  \hline
$M$ & $\partial M$ & Seed Metric & $deg \Pi^{0}$ & $\Pi^{0}$ onto ${\cal C}^{0}$ \\ \hline\hline
$B^{4}$ & $S^{3}$ & Poincar\'e & 1 & Yes \\ \hline
${\Bbb C}{\Bbb P}^{2} \setminus B^{4}$ & $S^{3}$ & AdS Taub-Bolt & 0 & No \\ \hline 
$S^{2}\times {\Bbb R}^{2}$ & $S^{2}\times S^{1}$ & AdS Schwarzschild & 0 & No \\ \hline 
${\Bbb R}^{3}\times S^{1}$ & $S^{2}\times S^{1}$ & Poincar\'e/${\Bbb Z}$ & 1 & Yes \\ \hline  
$E_{k}\rightarrow S^{2}$, $k \geq 2$ & $S^{3}/{\Bbb Z}_{k}$ & AdS Taub-Bolt & 1 & Yes \\ \hline 
$X_{k}$, $k \geq 2$ & $S^{3}/{\Bbb Z}_{k}$ & Self-dual CS metrics & 0 & No \\ \hline

\end{tabular}

\end{center}

\renewcommand{\arraystretch}{1}

\bigskip

  Here $(M, \partial M)$ is the given manifold with boundary and the existence of a 
seed metric implies ${\cal E} \neq \emptyset$. The degree is given for the component 
of ${\cal E}$ containing the seed metric. 

  The manifold $E_{k}$ is the ${\Bbb R}^{2}$ bundle over $S^{2}$ with Chern class $k$, 
while $X_{k}$ is a resolution of the orbifold ${\Bbb C}^{2}/{\Bbb Z}_{k}$ with $c_{1}(X_{k}) 
< 0$. The seed metric on $X_{k}$ is an element of the family of self-dual AH Einstein 
metrics on $X_{k}$ constructed by Calderbank-Singer, [17].

  Mazzeo-Pacard [43] have shown that if $M_{1}$ and $M_{2}$ admit an AH Einstein metric, 
then so does the boundary connected sum $\hat M = M_{1} \#_{b} M_{2}$; for $\hat M$ one 
has $\partial \hat M = \partial M_{1} \# \partial M_{2}$. Further, if ${\cal E}^{0}(M_{i}) 
\neq \emptyset$, then ${\cal E}^{0}(\hat M) \neq \emptyset$. It would be 
interesting to determine the degree of $\hat M$ in terms of the degree of 
each $M_{i}$.

\begin{remark} \label{r 3.4 (i).}
 {\rm  Let $M$ be any $(n+1)$ manifold with $\partial M = S^{n}.$ If 
$M \neq  B^{n+1}$, then $\Pi$ cannot be surjective onto ${\cal C}^{0}$; in 
particular when $n+1 = 4$, $deg \Pi^{0} = 0$. In fact the round metric 
$g_{+1}$ on $S^{n}$ cannot be in Im $\Pi$, for Theorem 3.3 implies that 
any such AH Einstein metric must be the hyperbolic metric on the ball. The 
same argument shows that the conformal class of the round product metric 
$S^{1}(\beta )\times S^{n}(1)$, for $\beta > \beta_{0} = 2\pi /(n(n-2))^{1/2}$, 
is not in Im $\Pi$ on any manifold $M$ with $\partial M = S^{1}\times S^{n}$ 
except ${\Bbb R}^{n}\times S^{1}$, where it is uniquely realized by the 
hyperbolic metric.

  In sum, one has the following examples of uniqueness results from Theorem 3.3. 
The Poincar\'e metric is the unique AH Einstein metric with boundary metric the 
round metric on $S^{n}$, while the AdS Schwarzschild and (quotient) Poincar\'e 
metric are unique among AH Einstein metrics with boundary the round product on 
$S^{n-1}\times S^{1}$. Similarly, the only manifold carrying an AH Einstein metric 
with boundary metric the round metric on $S^{3}/{\Bbb Z}_{k}$ is $E_{k}$ when 
$k \geq 2$, and it is uniquely realized by the AdS Taub-Bolt metric, 
cf. [32] for example. 

{\bf (ii).}
 By the same reasoning as above, any AH Einstein metric on $M$ with 
conformal infinity a flat metric on the $n$-torus $T^{n}$ is necessarily 
a twisted AdS toral black hole metric, as in Examples 2.2. It follows from a 
simple Wick rotation argument that the Lorentzian AdS soliton metric of 
Horowitz-Myers [37], is the unique static AdS metric, when the 
conformal infinity is compactified to a flat torus; see also [29], [10] for 
previous work on the uniqueness of the AdS solition. }
\end{remark}

\section{Role of $R \geq  0$.}
\setcounter{equation}{0}

 In this section, we discuss the role of the hypothesis that the 
boundary metric $\gamma $ on $\partial M$ has non-negative scalar 
curvature. This condition was first suggested by Witten in [47], who 
pointed out that the corresponding CFT is unstable when $R_{\gamma} <$ 
0, suggesting that the AdS/CFT correspondence may break down in this 
region. This result led shortly thereafter to the work of Witten-Yau [48], 
proving that $\partial M$ is necessarily connected when $\partial M$ 
carries a metric of positive scalar curvature, see also the work of 
Cai-Galloway [16] for a different proof which handles in addition the 
case of non-negative scalar curvature..

 In this section we give a very elementary proof of the Witten-Yau 
result. In fact the result is stronger in that it gives a definite 
bound on the distance of any point in $M$ to its boundary, in the 
geodesic compactification. This shows that the condition $R_{\gamma} >$ 
0 not only proves that $\partial M$ must be connected, but also 
prevents the formation of new, not necessarily conformally compact, 
boundary components in families of AH metrics. 

\medskip

 Let $g$ be an AH metric (not necessarily Einstein), on an 
$(n+1)$-manifold $M$, possibly with several boundary components; thus $g$ 
is merely assumed to be $C^{3}$ conformally compact. Given a fixed 
boundary component $\partial_{0}M,$ with associated boundary metric 
$\gamma ,$ let $\rho $ be the geodesic defining function defined by 
$(\partial_{0}M, \gamma ),$ so that if $\bar g = \rho^{2}g$ is the 
associated (partial) $C^{2}$ compactification of $(M, g)$, then $\rho (x) 
= dist_{\bar g}(x, \partial_{0}M).$ Observe that if $M$ has other 
boundary components, then these lie at infinite distance with respect 
to $\bar g$ to any point in $M$. Note also that since $g$ is $C^{2}$ 
conformally compact, $|Ric_{g}+ng| = O(\rho^{2}).$

\begin{theorem} \label{t 4.1.}
  Let $\bar g = \rho^{2}g$ be a partial geodesic compactification of an 
AH metric $g$ on M, satisfying
\begin{equation} \label{e4.1}
Ric_{g} + ng \geq  0  \ \ {\rm and} \ \   |Ric_{g}+ng| = o(\rho^{2}). 
\end{equation}
If $R_{\gamma} =$ const $> $ 0, then for all $x\in M,$
\begin{equation} \label{e4.2}
\rho^{2}(x) \leq  4n(n-1)/R_{\gamma}. 
\end{equation}
\end{theorem}

{\bf Proof:}
 Along the $\bar g$-geodesics normal to $\partial_{0}M$, one has the 
Riccati equation
\begin{equation} \label{e4.3}
\bar H'  + |\bar K|^{2} + \bar Ric(T,T) = 0, 
\end{equation}
where $T = \bar \nabla \rho$, $\bar K$ is the 2nd fundamental form of 
the level sets $S(\rho)$ of $\rho$, and $\bar H = tr \bar K$ is the mean 
curvature, with $\bar H' = \partial \bar H / \partial \rho$. Thus $\bar K = 
\bar D^{2}\rho$, $\bar H = \bar \Delta\rho$. Here and in the following, 
the computations are with respect to $\bar g$. Standard formulas for 
conformal changes of metric give
\begin{equation} \label{e4.4}
\bar Ric = -(n-1)\frac{\bar D^{2}\rho}{\rho} - 
\frac{\bar \Delta\rho}{\rho}\bar g + (Ric_{g}+ng) \geq  
-(n-1)\frac{\bar D^{2}\rho}{\rho} - \frac{\bar \Delta\rho}{\rho}\bar g. 
\end{equation}
Hence 
\begin{equation} \label{e4.5}
\bar R = -2n\frac{\bar \Delta\rho}{\rho} + (R+n(n+1))/\rho^{2} \geq  
-2n\frac{\bar \Delta\rho}{\rho}. 
\end{equation}
In particular, $\bar Ric(T,T) = -\frac{\bar \Delta\rho}{\rho} + 
(Ric_{g}+ng)(T,T)/\rho^{2}.$

 Dividing (4.3) by $\rho$ then gives
$$\frac{(\bar \Delta\rho )'}{\rho} -\frac{\bar \Delta\rho}{\rho^{2}} + 
(Ric_{g}+ng)(T,T)/\rho^{3} + \frac{|\bar D^{2}\rho|^{2}}{\rho} = 0, $$
so that
\begin{equation} \label{e4.6}
(\frac{\bar \Delta\rho}{\rho})'  = - \frac{|\bar D^{2}\rho|^{2}}{\rho} 
- (Ric_{g}+ng)(T,T)/\rho^{3}  \leq  - \frac{|\bar 
D^{2}\rho|^{2}}{\rho}. 
\end{equation}
By Cauchy-Schwartz, $|\bar D^{2}\rho|^{2} \geq  \frac{1}{n}(\bar 
\Delta\rho )^{2}$ and so, setting $\phi  = -\bar \Delta\rho /\rho $ one 
has
\begin{equation} \label{e4.7}
\phi'  \geq  \frac{1}{n}\rho\phi^{2}. 
\end{equation}
A simple computation using the Gauss equations at $\partial M$ together 
with the fact that $A = 0$ at $\partial M$ implies that
\begin{equation} \label{e4.8}
(n-1)\phi (0) = \tfrac{1}{2}R_{\gamma} + \lim_{\rho\rightarrow 
0}\frac{1}{\rho^{2}}[(Ric_{g}+ng)(T,T) - \tfrac{1}{2}(R + n(n+1))]. 
\end{equation}
The hypothesis (4.1) implies that the limit is 0, and hence $\phi (0) 
>$ 0. A simple integration then gives
\begin{equation} \label{e}
\rho^{2} \leq  2n/\phi (0), 
\end{equation}
which gives (4.2).
{\endproof}

 Note that if $\gamma\in [\gamma]$ has non-negative scalar curvature, 
then there exists a representative $\bar \gamma \in [\gamma]$ with 
constant non-negative scalar curvature, by the solution of the Yamabe 
problem.

 As noted above, the estimate (4.2) immediately implies that $\partial 
M$ is connected, since any other boundary component would have to lie 
at infinite $\rho$-distance to $\partial_{0}M.$ Similarly, simple 
topological arguments then imply
$$\pi_{1}(\partial M) \rightarrow  \pi_{1}(M) \rightarrow  0. $$

 More importantly for our purposes, the estimate (4.2) also immediately 
implies that AH Einstein metrics with cusps, as described in \S 3, 
cannot form as the limit of sequences of AH Einstein metrics with 
boundary metrics of uniformly positive scalar curvature. This explains 
then the role of ${\cal C}^{0}$ in Theorem 3.1. A straightforward 
extension of the method above (based on the Cheeger-Gromoll splitting 
theorem) shows also that if $R_{\gamma} = 0$ and $\rho$ is unbounded 
on $(M, g)$, then $(M, g)$ is isometric to 
$$g = ds^{2} + e^{2s}g_{N^{n}}, $$
where $g_{N^{n}}$ is Ricci-flat, cf. [16]. Thus when $n = 3$, $N$ must be 
flat so a finite cover of $(M, g)$ is isometric to a hyperbolic cusp 
metric (2.8). In particular, this can only happen if $(\partial M, 
\gamma)$ is flat.

\begin{remark} \label{r 4.2.}
 {\rm  Theorem 4.1 was proved in [4]. The proof is included here 
partly for completeness, and partly because the Lorentzian version of 
this result will be used in \S 6. 

  An elementary consequence of Theorem 4.1 is that a geometrically finite 
hyperbolic manifold with conformal infinity satisfying $R_{\gamma} > 0$ has 
no parabolic ends.}
\end{remark}

\section{Self-Duality.}
\setcounter{equation}{0}

 The analysis in the previous sections describes the beginnings of a 
well-defined existence theory for the Einstein-Dirichlet problem, at least 
in $4$-dimensions. From the point of view of the AdS/CFT correspondence, one 
would like however much more detailed information about the correspondence 
(1.9) of the Dirichlet and Neumann boundary data.

 Again, restricting to dimension 4, a good deal more can be said, using 
the splitting of the curvature tensor into self-dual and 
anti-self-dual parts. Thus, let $M = M^{4}$ be an oriented 4-manifold 
with boundary. As first noticed by Hitchin [35], given any $C^{2}$ 
conformally compact metric $g$ on $M$, (not necessarily Einstein), a 
simple calculation using the Atiyah-Patodi-Singer index theorem gives
\begin{equation} \label{e5.1}
\frac{1}{12\pi^{2}}\int_{M}|W^{+}|^{2} - |W^{-}|^{2} = \sigma (M) - 
\eta (\gamma ), 
\end{equation}
where $W^{\pm}$ are the self-dual and anti-self-dual parts of the 
Weyl curvature, $\sigma (M)$ is the signature of M, and $\eta (\gamma 
)$ is the eta invariant of the conformal infinity $(\partial M, \gamma 
);$ note that (5.1) is conformally invariant. Combining this with the 
formula (1.20) for the renormalized action of an AH Einstein metric 
gives the formula
\begin{equation} \label{e5.2}
I^{ren} \leq  8\pi^{2}\chi (M) -12\pi^{2}|\sigma (M) - \eta (\gamma )|, 
\end{equation}
with equality if and only if $g$ is self-dual, (or anti-self-dual).

 Hence if $(M, g)$ is self-dual and Einstein, 
\begin{equation} \label{e5.3}
I^{ren} = 12\pi^{2}\eta (\gamma ) + (8\pi^{2}\chi (M) - 12\pi^{2}\sigma 
(M)), 
\end{equation}
while $I^{ren} = -12\pi^{2}\eta (\gamma ) + (8\pi^{2}\chi (M) + 
12\pi^{2}\sigma (M)),$ if $(M, g)$ is anti-self-dual Einstein.

 This leads to the following result:

\begin{theorem} \label{t 5.1.}
  Let ${\cal E}_{sd}$ be the moduli space of self-dual AH Einstein 
metrics on a 4-manifold $M$. Then $I^{ren}: {\cal E}_{sd} \rightarrow  
{\Bbb R}$ is given by
\begin{equation} \label{e5.4}
I^{ren} = 12\pi^{2}\eta  + c_{M}, 
\end{equation}
where $c_{M} = 8\pi^{2}\chi (M)$ - $12\pi^{2}\sigma (M)$ is topological.

 The stress-energy tensor $g_{(3)}$ at a self-dual Einstein metric is 
given by
\begin{equation} \label{e5.5}
dI^{ren} = -\tfrac{3}{2}g_{(3)} = 12\pi^{2}d\eta  =  -\tfrac{1}{2}*dRic. 
\end{equation}
\end{theorem}

{\bf Proof:}
 Both (5.4) and (5.5) follow immediately from (5.3). In (5.5), $Ric$ is 
viewed as a 1-form with values in $T(\partial M)$ and $*$ is the Hodge 
$*$-operator. The formula for $d\eta$ comes from the original work of 
Chern-Simons [18], cf. also [3] for more details.
{\endproof}

 Similarly on the moduli space of anti-self-dual AH Einstein metrics 
${\cal E}_{ads}$, $I^{ren} = -12\pi^{2}\eta  + c_{M}'$, and $dI^{ren} = 
-12\pi^{2}d\eta  = \frac{1}{2}*dRic$.

 Thus, on the moduli spaces ${\cal E}_{sd}$ or ${\cal E}_{asd}$, the 
renormalized action is explicitly computable from the {\it  global}  
geometry of the boundary metric $(\partial M, \gamma )$, while the 
stress-energy is {\it  locally computable}  on $(\partial M, \gamma)$. 

\begin{remark} \label{r 5.2.}
 {\rm  This exact identification of the renormalized gravitational action 
and its stress-energy tensor on ${\cal E}_{sd}$ or ${\cal E}_{asd}$ 
appears to closely resemble the identification of the renormalized 
action with the Liouville action on the boundary in dimensions $2+1$, 
cf. [38], [20]. It would be very interesting to pursue this analogy further. }
\end{remark}

 The discussion above suggests that a self-dual (or anti-self-dual) 
AH Einstein metric should be uniquely determined by its boundary metric 
$\gamma .$ This is true at least for real-analytic boundary data, 
$\gamma\in C^{\omega},$ as proved by LeBrun [40], using twistor 
methods. A more elementary proof, using just the Cauchy-Kovalewsky 
theorem, was given recently in [6]. This uniqueness implies that the 
topology of the bulk manifold $M$ is determined by the boundary data 
$(\partial M, \gamma)$, up to covering spaces, i.e. any two self-dual 
AH Einstein metrics $(M_{1},g_{1}), (M_{2}, g_{2})$ with the same 
boundary data are locally isometric; in particular they are isometric 
in some covering space.

 Similarly, in analogy to the discussion following (1.9), [40] or [6] 
imply that any metric $\gamma\in C^{\omega}(\partial M)$ is the 
boundary metric of a self-dual or anti-self-dual AH Einstein metric 
$g$, defined on a thickening $\partial M\times [0,\varepsilon )$ of $\partial 
M.$ Of course, for general $\gamma ,$ this metric will not extend to a 
smooth Einstein metric on a compact manifold $M$ with boundary 
$\partial M.$

 For example, consider the boundary $S^{2}\times S^{1}$, with boundary metric 
the round conformally flat product metric. This bounds a self-dual AH 
Einstein metric in a neighborhood of $S^{2}\times S^{1}$, and uniqueness 
implies that this metric must be the hyperbolic metric on 
$B^{3}\times S^{1}$, given by $H^{4}(-1)/{\Bbb Z}$. The AdS Schwarzschild 
metric $S^{2}\times {\Bbb R}^{2}$ is thus of course not self-dual. 

\medskip

 It would interesting to know if the moduli space ${\cal E}_{sd}$ has 
the structure of an infinite dimensional manifold, as is the case with 
${\cal E} $ itself. Noteworthy in this respect is a result of Biquard [14] 
that in a neighborhood of the hyperbolic metric on $B^{4}$, the 
spaces ${\cal E}_{sd}$ and ${\cal E}_{asd}$ are smooth infinite 
dimensional manifolds, which intersect transversally at the hyperbolic 
metric $g_{-1}$. In particular, any metric $g\in{\cal E} $ near 
$g_{-1}$ can be uniquely written as a sum $g = g^{+} + g_{-1} + g^{-}$, 
where $g^{+} + g_{-1}$ is self-dual and $g_{-1} + g^{-}$ is anti-self-dual. 
It seems that this result should be useful in understanding the 
Dirichlet-Neumann correspondence (1.9) near $g_{-1}$. 

 There are a number of interesting and explicit or semi-explicit 
examples of self-dual AH Einstein metrics. Thus, the AdS Taub-NUT [32] 
or Pedersen metrics [44] are self-dual on $B^{4},$ with boundary metric 
a Berger sphere, (the 3-sphere squashed along the $S^{1}$ fibers of the 
Hopf fibration). More generally, Hitchin [36] has constructed self-dual AH 
Einstein metrics on $B^{4}$ with conformal infinity any left-invariant 
metric on $SU(2) = S^{3}$. In addition, LeBrun [41] has proved the 
existence of an infinite dimensional family of self-dual AH Einstein 
metrics on $B^{4}$. More recently, Calderbank and Singer [17] have 
constructed families of self-dual AH Einstein metrics on resolutions of 
orbifolds ${\Bbb C}^{2}/{\Bbb Z}_{k}$ having negative Chern class. This 
gives 4-manifolds with arbitrarily large Betti number $b_{2}$ for which 
${\cal E}_{sd} \neq \emptyset$. 

\section{Continuation to de Sitter and self-similar vacuum space-times.}
\setcounter{equation}{0}

 In this section we discuss the continuation of AH Einstein metrics to 
de Sitter-type Lorentz metrics, and the possibility of constructing global 
self-similar vacuum solutions of the Einstein equations in higher 
dimensions. The basic model for this picture is the decomposition of 
Minkowski space-time $({\Bbb R}^{4}, \eta )$ into foliations by 
hyperbolic metrics in the interior of the past and future light cones 
of a point \{0\}, and foliations by deSitter metrics in the exterior of 
the light cone. These foliations are of course given as the level sets 
of the distance function to \{0\}. Some aspects of this work have 
previously appeared in [8], see also [46] for a formal treatment of 
some of these issues. 

\medskip

 Let $M$ be any compact $(n+1)$ manifold with boundary $\partial M$ and 
let $g$ be any AH Einstein metric on $M$, with boundary metric 
$(\partial M, \gamma)$, with respect to a geodesic defining function $\rho$. 
As is well-known, and observed by Fefferman-Graham [27] in their original 
work on the subject, $(M, g)$ then generates a vacuum solution to the 
Einstein equations on ${\cal M}  = {\Bbb R}^{+}\times M$ given by 
\begin{equation} \label{e6.1}
{\frak g} = -d\tau^{2} + \tau^{2}g, 
\end{equation}
where $\tau\in (0,\infty)$. This is a Lorentzian cone metric on the 
Riemannian metric $g$, and satisfies the vacuum equations
\begin{equation} \label{e6.2}
Ric_{\frak g} = 0. 
\end{equation}
At least in a neighborhood of ${\Bbb R}^{+}\times \partial M$, by (1.3), the 
metric (6.1) may be rewritten in the form
\begin{equation} \label{e6.3}
{\frak g} = -d\tau^{2} + (\frac{\tau}{\rho})^{2}(d\rho^{2} + g_{\rho}). 
\end{equation}

 The space-time $({\cal M}, {\frak g})$ is globally hyperbolic, with Cauchy 
surface given by $M$ and Cauchy data $(g, K) = (g, g)$. The time 
evolution with respect to the time parameter $\tau$ is given by 
trivial rescalings of the time 1 spatial metric $(M, g)$. When $(M, g) = 
({\Bbb R}^{3}, g_{-1})$ is the Poincar\'e metric on the 3-ball, 
$({\cal M}, {\frak g})$ is the interior of the past (or future) light cone 
of a point \{0\} in Minkowski space-time ${\Bbb R}^{4}$, (also called the Milne 
universe). Similarly, for any $(M, g)$ with $g\in{\cal E}$, the vacuum 
solution $({\cal M}, {\frak g})$ is the interior of the past (or future) light 
cone $(H, \gamma_{0})$, where $H = {\Bbb R}^{+}\times \partial M$ and the 
degenerate metric $\gamma_{0}$ on $H$ is given by
$$\gamma_{0} = v^{2}\gamma , $$
with $v\in {\Bbb R}^{+}$ given by $v = \tau /\rho$. Thus, $(H, \gamma_{0})$ 
is the smooth Cauchy horizon for the space-time $({\cal M}, {\frak g})$. To 
be definite, we choose $({\cal M}, {\frak g})$ to be the interior of the past 
light cone of the vertex $\{0\} = \{v=0\}$, and will later set 
$({\cal M}, {\frak g}) = ({\cal M}^{-}, {\frak g}^{-})$ and $H = H^{-}$.

\medskip

 In general, the metric ${\frak g}$ is not $C^{\infty}$ up to the Cauchy 
horizon $H$. Thus, under the change of variables $(\tau , \rho) 
\rightarrow  (v, x)$, with $\rho  = \sqrt{x}$, (6.3) becomes
\begin{equation} \label{e6.4}
{\frak g} = -xdv^{2} - vdvdx + v^{2}g_{\sqrt{x}}, 
\end{equation}
and for $g_{\sqrt{x}}$ one has the Fefferman-Graham expansion
\begin{equation} \label{e6.5}
g_{\sqrt{x}} = \gamma  + xg_{(2)} + ... + x^{n/2}g_{(n)} + 
\tfrac{1}{2}x^{n/2}\log x \ h + ... . 
\end{equation}
Hence, if $n$ is odd, the metric ${\frak g}$ is $C^{n/2}$ up to the horizon $H$, 
while if $n$ is even, ${\frak g}$ is $C^{n/2-\varepsilon}$ up to $H$. This 
degree of smoothness cannot be improved by passing to other coordinate 
systems; only in very rare instances where $g_{(n)} = 0$ when $n$ is 
odd, or $h = 0$ when $n$ is even, will $g$ be $C^{\infty}$ up to $H$.

\medskip

 Suppose first that $n = 3$ and let $\gamma$ be a real-analytic metric 
on $\partial M$. Then by [6], the compactification $\bar g = 
\rho^{2}g$ of $(M, g)$ is also real-analytic on $\bar M$, and so the 
Fefferman-Graham expansion (1.4) converges to $g_{\rho}$. Thus, the 
curve $g_{\rho}$ can be extended past $\rho = 0$ to purely imaginary 
values of $\rho$. This corresponds to replacing  $\sqrt{x}$, $x > 0$, by  
$-\sqrt{|x|}$ , $x <  0$, (i.e. $\rho  \rightarrow  i\rho$), and so gives 
the curve
\begin{equation} \label{e6.6}
g_{\rho}^{ext} = \gamma  - \rho^{2}g_{(2)} - \rho^{3}g_{(3)} + 
\rho^{4}g_{(4)} + .... , 
\end{equation}
obtained from the expansion for $g_{\rho}$ by replacing $\rho $ by 
$i\rho $ and dropping the $i$ coefficients. Of course one could also 
continue the Fefferman-Graham expansion into the region $\rho  <$ 0; 
this would give an AH Riemannian Einstein metric on ``the other side'' of 
$\partial M$, defined at least in some neighborhood of $\partial M$. 
However, this extension will not be of concern here.

 Thus, although the metric $({\cal M}, {\frak g})$ is only $C^{3/2}$ at $H$, 
it extends via (6.6) and (6.4) across the horizon $H$ into the exterior of 
the light cone. Returning to the original variables $(\tau , \rho)$ in 
(6.3) then gives the metric
\begin{equation} \label{e6.7}
{\frak g}^{ext} = d\tau^{2} + (\frac{\tau}{\rho})^{2}(-d\rho^{2} + 
g_{\rho}^{ext}). 
\end{equation}
Formally, this is obtained from $g = g^{-}$ by the replacement $\tau  
\rightarrow  i\tau$, $\rho  \rightarrow  i\rho$, interchanging a 
spatial and time direction. This gives an extension of the metric 
${\frak g}$ into a region ${\cal M}^{ext}$ exterior to the light cone $H$, 
defined for all $\tau\in (0,\infty)$, $\rho\in [0,\varepsilon)$, for some 
$\varepsilon > 0$. The metric ${\frak g}^{ext}$ is $C^{\omega}$ where $\rho  
\neq 0$, but is only $C^{3/2}$ up to $H$ where $\rho = 0$. 

 The metric ${\frak g}^{ext}$ is also a Lorentzian cone metric, now however 
with a space-like self-similarity in $\tau$ in place of the previous 
time-like self-similarity. The slices $\tau  = const$ are all 
homothetic, and are Lorentzian metrics of the form
\begin{equation} \label{e6.8}
{\frak g}^{dS} = (\frac{1}{\rho})^{2}(-d\rho^{2} + g_{\rho}^{ext}). 
\end{equation}
The metric ${\frak g}^{dS}$ is a solution to the vacuum Einstein equations 
with a positive cosmological constant $\Lambda  = \frac{1}{2}n(n-1)$, 
i.e. when $n = 3$,
\begin{equation} \label{e6.9}
Ric_{{\frak g}^{dS}} = 3{\frak g}^{dS}, 
\end{equation}
and so ${\frak g}^{dS}$ is a deSitter-type (dS) space-time, (just as the 
initial metric $g$ is of anti-deSitter type).

\medskip

 Exactly the same discussion holds in all dimensions. Thus, suppose 
again that $\gamma  \in  C^{\omega}(\partial M).$ As noted in \S 1, it 
follows from the recent regularity result of Chru\'sciel et al. [19], 
that the compactification $\bar g = \rho^{2}g$ is $C^{\infty}$ 
polyhomogeneous. Moreover, recent work of Kichenassamy [39] or Rendall 
[45] implies the Fefferman-Graham expansion (1.4) or (1.6) converges to 
$g_{\rho},$ in both cases $n$ even or $n$ odd. Hence, exactly the same 
arguments as above hold for any $n$, and give a dS-type Einstein metric 
of the form (2.8) satisfying 
\begin{equation} \label{e6.10}
Ric_{{\frak g}^{dS}} = n{\frak g}^{dS}, 
\end{equation}
with $g_{\rho}^{ext}$ given by
\begin{equation} \label{e6.11}
g_{\rho}^{ext} = \gamma  - \rho^{2}g_{(2)} - \rho^{3}g_{(3)} + ...  
\pm \rho^{n}g_{(n)}  \pm \rho^{n}\log \rho \ h + .. .  
\end{equation}
The terms $g_{(k)}$ are defined as in (1.8), where $T = \bar \nabla\rho$ is 
the future-directed unit vector.

 We summarize some of this discussion in the following result.

\begin{corollary} \label{c 6.1.}
  Let $\gamma $ be a real-analytic metric on an n-manifold $\partial M$, 
and $g_{(n)}$ a real-analytic symmetric bilinear form on $\partial M$ 
satisfying the constraint conditions (1.5) or (1.7). Then there is a 
1-1 correspondence between Riemannian AH Einstein metrics $g$ with boundary 
metric $\gamma$, and deSitter-type Lorentzian Einstein metrics ${\frak g}^{dS}$ 
with past (or future) boundary metric $\gamma$, given by (6.3)-(6.8).
\end{corollary}
{\endproof}

 This correspondence thus gives a rigorous form of ``Wick rotation'' 
between these types of Einstein metrics. The Fefferman-Graham expansion 
holds for Einstein metrics of any signature, (again as observed in [27]). 
In the correspondence between AH and dS Einstein metrics, one has
\begin{equation} \label{e6.12}
g_{(k)}^{AH} = \pm g_{(k)}^{dS}, 
\end{equation}
where $+$ occurs if $k \equiv  0, 1$ (mod 4), while $-$ occurs if $k \equiv 
2, 3$ (mod 4).

 In dimension 4, the formulas (1.20) and (1.19) for the renormalized 
action and its variation also have analogues for dS space-times. Thus, 
let $({\cal S},{\frak g})$ be a solution of (6.10) which is asymptotically 
simple, in that $({\cal S}, {\frak g})$ has a smooth past and future conformal 
infinity $({\cal I}^{-}, \gamma^{-})$ and $({\cal I}^{+}, \gamma^{+})$. In 
particular, ${\cal S}$ is geodesically complete and globally 
hyperbolic with compact Cauchy surface $\Sigma$, a 3-manifold 
diffomorphic to ${\cal I}^{-}$ and ${\cal I}^{+}$. In the following, we will 
forgo the exact determination of signs, which are best computed on an example; 
note that the Einstein-Hilbert action (1.11) is usually replaced by its 
negative for Lorentzian metrics. 

\begin{proposition} \label{p 6.2.}
  Let $({\cal S}, {\frak g})$ be a $4$-dimensional asymptotically simple vacuum 
dS space-time. Then
\begin{equation} \label{e6.13}
\pm I^{ren} = \int_{{\cal S}}|W|^{2}dV, 
\end{equation}
where $|W|^{2} = W_{ijkl}W^{ijkl}$ and
\begin{equation} \label{e6.14}
\pm dI^{ren} = g_{(3)}^{+} - g_{(3)}^{-}, 
\end{equation}
where the terms are taken with respect to the past unit normal.

 Since the metric $g$ is Lorentzian, note that $|W|^{2}$ is not apriori 
non-negative.
\end{proposition}

{\bf Proof:}
 The proof of (6.13) is exactly the same as the proof of (1.20) in [3], 
using the Lorentz version of the Chern-Gauss-Bonnet theorem, cf. [1] 
for example, in place of the Riemannian version. Since ${\cal S}  = 
{\Bbb R} \times \Sigma$ topologically, where $\Sigma$ is a closed 
3-manifold, $\chi ({\cal S} ) = 0$. Similarly, the proof of (1.19) in 
[3] holds equally well for Lorentzian metrics, and gives (6.14).
{\endproof}

 Returning to the discussion preceding Corollary 6.1, the extended 
metric ${\frak g}$ on the enlarged space ${\cal M}^{-}\cup{\cal M}^{ext}$ is 
still a solution to the vacuum Einstein equations, (with $\Lambda  = 0$). 
This is clear if $n > 4$, since the metric is everywhere at least 
$C^{5/2}$. For $n = 3, 4$, the metric is not $C^{2}$, but is easily 
verified to still be a weak solution of the vacuum Einstein equations, 
i.e. it satisfies the equations (6.2) distributionally.

 Now the initial AH Einstein metric $(M, g)$ is global. It is natural to 
ask if the dS metric ${\frak g}^{dS}$ is also global; the formula (6.7) is only 
defined for $\rho\in [0,\varepsilon)$, for some $\varepsilon > 0$. 
In general, the answer is no. In fact, the work in \S 4 carries over to 
this setting almost identically, and gives the following result, proved 
independently by the author (unpublished) and Andersson-Galloway [11].
\begin{proposition} \label{p 6.3.}
  Let $({\cal S}, {\frak g})$ be an $(n+1)$ dimensional globally hyperbolic 
space-time, with compact Cauchy surface $\Sigma$, which is $C^{3}$ 
conformally compact to the past, so that past conformal infinity 
$({\cal I}^{-}, \gamma)$ is $C^{3}$. Suppose $({\cal S}, {\frak g})$ 
satisfies the strong energy and decay conditions
\begin{equation} \label{e6.15}
(Ric_{{\frak g}} - n{\frak g})(T,T) \geq  0  \ \ {\rm and} \ \  
|(Ric_{{\frak g}}-n{\frak g})(T,T)| = o(\rho^{2}), 
\end{equation}
for $T$ time-like. Let $\gamma$ be a representative for $[\gamma]$ 
with constant scalar curvature $R_{\gamma}$. If $R_{\gamma} < 0$, then
\begin{equation} \label{e6.16}
\rho^{2}(x) \leq  4n(n-1)/|R_{\gamma}|, 
\end{equation}
where $\rho$ is the geodesic defining function associated to 
$({\cal I}^{-}, \gamma)$.

 In particular, any time-like geodesic in ${\cal S}$ is future 
incomplete, and no Cauchy surface $\Sigma_{\rho}$ exists, even 
partially, for $\rho^{2} > 4n(n-1)/|R_{\gamma}|$, so that ${\cal I}^{+} = 
\emptyset$. 
 \end{proposition}

{\bf Proof:}
 The proof is essentially identical to that of Theorem 4.1. Let $\bar {{\frak g}} 
= \rho^{2} {\frak g}$ be the $C^{2}$ geodesic compactification determined by the 
data $({\cal I}^{-}, \gamma)$; as before the computations below are 
with respect to $\bar {{\frak g}}$. The equation (4.3) holds for Lorentzian 
metrics, (where it is known as the Raychaudhuri equation). The vector 
field $T = \bar \nabla \rho$ is now a unit time-like vector field, so 
${\frak g}(T,T) = -1$. This has the implication that $H = -\bar \Delta\rho$ 
while
$$\bar Ric = -(n-1)\frac{\bar D^{2}\rho}{\rho} - 
\frac{\bar \Delta\rho}{\rho}\bar {{\frak g}} + (Ric_{{\frak g}}-n{\frak g}) \geq  
-(n-1)\frac{\bar D^{2}\rho}{\rho} - \frac{\bar \Delta\rho}{\rho}\bar {{\frak g}}, $$
and
$$\bar R = -2n\frac{\bar \Delta\rho}{\rho} + (R-n(n+1))/\rho^{2} \geq  
-2n\frac{\bar \Delta\rho}{\rho}. $$
In particular, $\bar Ric(T,T) = \frac{\bar \Delta\rho}{\rho} + 
(Ric_{g}-ng)(T,T)/\rho^{2}.$ The same arguments as in (4.4)-(4.7) then 
give
\begin{equation} \label{e6.17}
\phi'  \leq  -\frac{1}{n}\rho\phi^{2}, 
\end{equation}
where again $\phi  = -\bar \Delta\rho /\rho .$ The formula (4.8) also 
holds, and hence $R_{\gamma} <$ 0 implies $\phi (0) <$ 0. Integrating 
(6.17) as before gives (6.16).
{\endproof}

 It is straightforward to extend Proposition 6.3 to the situation where 
$R_{\gamma} = 0$. As in the case of Theorem 4.1, $({\cal S}, {\frak g})$ 
has ${\cal I}^{+} = \emptyset$ and no time-like geodesic is future-complete 
unless $({\cal S}, {\frak g})$ is isometric to 
$${\frak g} = -dt^{2} + e^{2t}g_{N^{n}}, $$
where $g_{N^{n}}$ is Ricci-flat, cf. [11] for further details. 

 Proposition 6.3 implies that dS space-times $({\cal S}, {\frak g})$ 
satisfying the strong energy and decay conditions (6.15) cannot be 
geodesically complete if $R_{\gamma} <$ 0, and have at most one 
component of conformal infinity. This exhibits the role of the 
hypothesis $R_{\gamma} >$ 0 in a more drastic way than the AH case.

\medskip

 We are interested in understanding when the vacuum dS metric ${\frak g}^{dS}$ 
constructed in (6.8) is also complete to the future, and has a smooth 
future conformal infinity ${\cal I}^{+}.$ In dimensions $n+1 >$ 4, it 
is an interesting open problem to find sufficient conditions 
guaranteeing the existence of complete asymptotically simple vacuum dS 
space-times. However, in dimension 4, a basic result of H. Friedrich 
does give global existence of dS vacuum solutions, for small 
perturbations of the exact deSitter metric.

\begin{theorem} \label{t 6.4.}
  {\rm [28]} Let $\gamma$ be a smooth metric on $S^{3}$ and $\sigma$ be a 
smooth transverse-traceless symmetric bilinear form on $S^{3}$. 
Suppose that $\gamma$ is sufficiently close to the round metric 
$g_{+1}$ on $S^{3}$ and $|\sigma|$ is sufficiently small, (measured 
with respect to $\gamma$). Then there exists a unique asymptotically 
simple vacuum dS space-time $({\cal S}, {\frak g})$, ${\cal S}  = S^{3}\times 
{\Bbb R}$ with smooth conformal compactification $\bar {\cal S} = 
{\cal S}  \cup  {\cal I}^{-}\cup{\cal I}^{+}$ for which the Fefferman-Graham 
expansion satisfies $g_{(0)} = \gamma$ and $g_{(3)} = \sigma$ on 
${\cal I}^{-}.$ If $\gamma$ and $\sigma$ are $C^{\omega}$, then 
$({\cal S}, {\frak g})$ is $C^{\omega}$ conformally compact.
\end{theorem}

 Note that, in contrast to the AH or AdS situation, $g_{(0)}$ and 
$g_{(3)}$ are freely specifiable on ${\cal I}^{-},$ subject to 
smallness conditions. An alternate version of the result should allow one 
to freely specify $g_{(0)}$ on ${\cal I}^{+}$ and ${\cal I}^{-},$ provided 
they are both close to the round metric on $S^{3}$; this remains to be 
proved however.  

\begin{remark} \label{r 6.5.}
 {\rm  This result is an exact analogue of the result of Graham-Lee [30] 
on AH Einstein perturbations of the Poincar\'e metric on the ball 
$B^{n+1};$ (since Friedrich's result predates that of Graham-Lee, the 
opposite statement is more accurate). It would be very interesting if a 
higher dimensional analogue of Friedrich's result could be proved, as 
in the Graham-Lee theorem. }
\end{remark}

 We may now apply this result to the ``initial'' AH Einstein metric $(M, g)$, 
$g = g^{-}$. Thus, on $M = B^{4}$, let $g^{-}$ be an AH Einstein 
metric with $C^{\omega}$ boundary metric $\gamma^{-}$ close to the 
round metric $\gamma_{+1}$ on $S^{3},$ (so $g^{-}$ is close to the 
Poincare metric on $B^{4}$). The metric $g^{-}$ determines the terms 
$\gamma  = g_{(0)} = g_{(0)}^{AH}$ and $g_{(3)}^{AH}$ in the 
Fefferman-Graham expansion. Let ${\frak g}^{dS}$ be the unique vacuum dS 
solution given by Friedrich's theorem satisfying
\begin{equation} \label{e6.18}
(g_{(0)}^{dS})_{{\cal I}^{-}} = g_{(0)}^{AH}, \ \ {\rm and} \ \   
(g_{(3)}^{dS})_{{\cal I}^{-}} = - g_{(3)}^{AH} , 
\end{equation}
where $g_{(3)}^{dS}$ is defined as in (1.8) with respect to the future normal 
$T = \bar \nabla \rho$. Thus, the stress-energy tensors of 
$g^{-}$ and ${\frak g}^{dS}$ cancel at ${\cal I}^{-}$.

 The vacuum dS solution ${\frak g}^{dS}$ is globally defined, and has a 
$C^{\omega}$ compactification to ${\cal I}^{-}$ and ${\cal I}^{+}$. Let 
$(g_{(0)}^{dS})_{{\cal I}^{+}}$ and $(g_{(3)}^{dS})_{{\cal I}^{+}}$ be 
the boundary metric and stress-energy tensor of ${\frak g}^{dS}$ at future 
conformal infinity ${\cal I}^{+}.$ Then $(g_{(0)}^{dS})_{{\cal I}^{+}}$ 
is close to the round metric $\gamma_{+1}$ on $S^{3},$ while 
$(g_{(3)}^{dS})_{{\cal I}^{+}}$ is close to 0, and both are 
real-analytic. By the Graham-Lee theorem [30], there is an AH Einstein 
metric $g^{+}$ on $B^{4}$ with boundary metric $\gamma_{+} = 
(g_{(0)}^{dS})_{{\cal I}^{+}},$ and by boundary regularity [6], 
$g^{+}$ has a real-analytic compactification.

 Thus, we have constructed a global $4+1$ dimensional space-time 
\begin{equation} \label{e6.19}
({\cal M}, {\frak g}) = ({\cal M}^{-}, g^{-})\cup ({\cal M}^{ext}, 
{\frak g}^{ext})\cup ({\cal M}^{+}, g^{+}). 
\end{equation}
This space-time is globally self-similar, with
\begin{equation} \label{e6.20}
{\cal L}_{\nabla\tau}{\frak g} = 2{\frak g}, 
\end{equation}
with a singularity at the vertex \{0\}. The metric ${\frak g}$ is 
$C^{\omega}$ off the null cone $H = H^{-}\cup H^{+}$ and is $C^{3/2}$ 
across the null-cone away from \{0\}. 

 In general however, it is not clear if $({\cal M}, {\frak g})$ is a vacuum 
space-time. The stress-energy tensor $g_{(3)}^{+}$ of $g^{+}$ is 
globally determined by the boundary metric $\gamma^{+},$ and there is 
no apriori reason that one should have
\begin{equation} \label{e6.21}
g_{(3)}^{+}= -(g_{(3)}^{dS})_{{\cal I}^{+}}, 
\end{equation}
as given by construction at ${\cal I}^{-}.$ Thus, there may be an 
effective stress-energy of the gravitational field along the future 
light cone $H^{+}$ of \{0\}. Of course the vacuum equations (6.9) are 
satisfied everywhere off $H^{+}.$

\medskip

 It is of interest to  understand if there exist non-trivial situations 
where (6.21) does hold, or to prove that it cannot hold. If (6.21) holds, then 
$({\cal M}, {\frak g})$ is a globally defined self-similar vacuum solution, with 
an isolated (naked) singularity at $\{0\}$. Of course in $3+1$ dimensions the 
only such space-time is empty Minkowski space $({\Bbb R}^{4}, \eta)$.

 We examine this issue on a particular family of examples.

{\bf Example 6.6.}
 Let $g^{-}$ be the AdS Taub-NUT metric on $B^{4}$, cf. [32] for example, 
(also called the Pedersen metric [44]), given by
\begin{equation} \label{e6.22}
g^{-} = \frac{E(r^{2}-1)}{F(r)}dr^{2} + 
\frac{EF(r)}{(r^{2}-1)}\theta_{1}^{2} + 
\frac{E(r^{2}-1)}{4}g_{S^{2}(1)}, 
\end{equation}
where $E \in  (0,\infty )$ is any constant, $r \geq $ 1, and 
\begin{equation} \label{e6.23}
F(r) = Er^{4} + (4-6E)r^{2} + (8E-8)r + 4-3E. 
\end{equation}
The length of the $S^{1}$ parametrized by $\theta_{1}$ is $2\pi$. 
This metric is self-dual Einstein and has conformal 
infinity $\gamma^{-}$ given by the Berger (or squashed) sphere with $S^{1}$ 
fibers of length $\beta = 2\pi E^{1/2}$ over $S^{2}(1)$. Clearly $\gamma^{-}$ 
is $C^{\omega}$, as is the geodesic compactification with boundary metric 
$\gamma^{-}$. Since $g^{-}$ is self-dual, the stress-energy tensor $g_{(3)}$ 
is given by (5.5). When $E = 1$, $g^{-}$ is the Poincar\'e metric.

 The deSitter continuation of $g^{-}$ is the dS Taub-NUT metric on 
${\Bbb R} \times S^{3}$, cf. [15] for instance, given by
\begin{equation} \label{e6.24}
{\frak g}^{dS} = -\frac{E(\tau^{2}+1)}{A(\tau )}d\tau^{2} + \frac{EA(\tau 
)}{(\tau^{2}+1)}\theta_{1}^{2} + \frac{E(\tau^{2}+1)}{4}g_{S^{2}(1)}, 
\end{equation}
where $\tau\in (-\infty ,\infty)$ and
\begin{equation} \label{e6.25}
A(\tau ) = E\tau^{4} - (4-6E)\tau^{2} - (8E-8)\tau  + 4-3E. 
\end{equation}
Again when $E = 1$, ${\frak g}^{dS}$ is the (exact) deSitter metric. For 
${\frak g}^{dS}$ to be complete and globally hyperbolic, without singularities, 
one needs $A(\tau ) > 0$, for all $\tau$. A lengthy but straightforward 
calculation shows this is the case exactly when
\begin{equation} \label{e6.26}
E \in  (\frac{2}{3},\frac{1}{3}(2 + \sqrt{3})). 
\end{equation}

 Suppose then $E$ satisfies (6.26). By construction, the metrics 
$g^{-}$ and ${\frak g}^{dS}$ satisfy (6.18). Observe from the explicit form of 
(6.24) that
\begin{equation} \label{e6.27}
\gamma^{-} = \gamma^{+}. 
\end{equation}
Thus, even though ${\frak g}^{dS}$ is not time-symmetric when $E \neq 1$, 
there is no gravitational scattering from past to future conformal 
infinity, in the sense that ${\cal I}^{-}$ is isometric to ${\cal I}^{+}$. 
However, further computation shows that (6.21) does not hold; instead one 
has $g_{(3)}^{+} = (g_{(3)}^{dS})_{{\cal I}^{+}}$, so that the full metric 
${\frak g}$ has an effective stress-energy tensor on the future null cone 
of $\{0\}$. We note however that there is an AH Taub-NUT metric satisfying 
(6.21) which has an isolated conical (nut) singularity at the origin of 
$B^{4}$; in this case, the formula (6.23) is replaced by a more general 
formula allowing two independent parameters, the mass and nut charge, 
in place of the one parameter $E$, cf. [24], [25]. If one fills in $H^{+}$ 
with such a metric, then the effective stress-energy tensor of $({\cal M}, 
{\frak g})$ is located on the future world line of the singularity 
$\{0\}$. 

\medskip

 Returning to the discussion of the de Sitter metrics (6.24), at the extremal 
values $E_{-} = \frac{2}{3}$ and $E_{+} = \frac{1}{3}(2 + \sqrt{3})$, the 
metric ${\frak g}^{dS}$ is still complete and globally hyperbolic. 
However, at these values, $g^{dS}$ is not in the space of metrics with smooth 
${\cal I}^{+}$ and ${\cal I}^{-}$; instead, it is in the boundary of this space. 
To explain this, let $\tau_{-} = -1$, and $\tau_{+} = 2 - \sqrt{3}$. Then at 
$E_{\pm}$, $A(\tau) \geq 0$, with $A(\tau) = 0$ exactly at $\tau_{\pm}$. Each  
metric $g^{dS}$ breaks up into a pair of complete, globally hyperbolic metrics, 
$g_{p}^{dS}$ and $g_{f}^{dS}$ parametrized on $(-\infty, \tau_{\pm})$ and 
$(\tau_{\pm}, \infty)$ respectively. The metric $g_{p}^{dS}$ has a smooth past 
conformal infinity ${\cal I}^{-}$, but ${\cal I}^{+} = \emptyset$, while 
$g_{f}^{dS}$ has a smooth future conformal infinity ${\cal I}^{+}$ but 
${\cal I}^{-} = \emptyset$. These metrics correspond to degenerate black hole 
metrics, and are analogous, in a dual sense, to the situation in Remark 2.3. 

  For $E$ outside the range (6.26), the function $A(\tau)$ changes sign, and 
the metrics $g^{dS}$ develop closed time-like curves, as with the behavior of 
the $\Lambda = 0$ Taub-NUT metrics

\medskip

 It is also straightforward to compute that $R_{\gamma^{-}} > 0$ 
exactly for $E$ in the range $E \in (0, 4)$. This shows that the 
converse of Proposition 6.3 does not hold, i.e. the condition 
$R_{\gamma} > 0$ is not sufficient to imply that a vacuum dS solution 
with smooth ${\cal I}^{-}$ is complete to the future.

\medskip

 It would be very interesting to generalize this example. For instance, 
can the same construction of globally self-similar almost-vacuum solutions be 
carried out for general AdS and dS Bianchi IX space-times, which have conformal 
infinity a general $SU(2)$ invariant metric on $S^{3}$? The AdS Bianchi IX metrics 
are self-dual, and have been described in detail by Hitchin [36]. Is the relation 
(6.27), related to self-duality?

\bibliographystyle{plain}

\bigskip
\begin{center}
March/June, 2004
\end{center}

\medskip
\noindent
\address{Department of Mathematics\\
S.U.N.Y. at Stony Brook\\
Stony Brook, NY 11794-3651}\\
\email{anderson@@math.sunysb.edu}

\end{document}